\documentclass[twocolumn]{openjournal}
\usepackage{natbib}
\usepackage{graphicx}
\usepackage{graphicx,amsmath,amssymb,amstext}
\usepackage{amsbsy,amsfonts,amsthm,color}
\usepackage{caption}
\usepackage{subcaption}
\usepackage{xcolor}
\usepackage{graphicx}
\usepackage{lipsum}
\usepackage{bm}
\usepackage{cancel}
\usepackage{float}
\usepackage{hyperref}
\hypersetup{colorlinks,linkcolor=ourblue,citecolor=ourblue,urlcolor=ourblue}
\usepackage{xcolor}
\definecolor{ourblue}{HTML}{118AB2}

\newcommand{\mvec}[1]{\bm{#1}}

\newcommand{\myvec}[1]{\mvec{#1}}

\DeclareUnicodeCharacter{2212}{-}

\usepackage{comment}

\usepackage{lineno}

\begin{document}

\journalinfo{The Open Journal of Astrophysics}
\shorttitle{\textrm{Inferring the Hubble Constant Using Simulated Strongly Lensed Supernovae and Neural Network Ensembles}}

\title{Inferring the Hubble Constant Using Simulated Strongly Lensed Supernovae and Neural Network Ensembles}

\author{Gon\c{c}alo Gon\c{c}alves$^{1,2}$}
\author{Nikki Arendse$^{2}$}
\author{Doogesh Kodi Ramanah$^{3}$}
\author{Rados{\l}aw Wojtak$^{3}$}

\affiliation{}
\affiliation{$^1$ Department of Physics, University of Porto, Rua do Campo Alegre, 4169-007 Porto, Portugal}
\affiliation{$^2$ Oskar Klein Centre, Department of Physics, Stockholm University, SE-106 91 Stockholm, Sweden}
\affiliation{$^3$ DARK, Niels Bohr Institute, University of Copenhagen, Jagtvej 128, 2200 Copenhagen, Denmark}

\email{}

\begin{abstract}

Strongly lensed supernovae are a promising new probe to obtain independent measurements of the Hubble constant (${H_0}$).
In this work, we employ simulated gravitationally lensed Type Ia supernovae (glSNe Ia) to train our machine learning (ML) pipeline to constrain $H_0$. 
We simulate image time-series of glSNIa, as observed with the upcoming Nancy Grace Roman Space Telescope, that we employ for training an ensemble of five convolutional neural networks (CNNs). 
The outputs of this ensemble network are combined with a simulation-based inference (SBI) framework to quantify the uncertainties on the network predictions and infer full posteriors for the $H_0$ estimates. We illustrate that the combination of multiple glSN systems enhances constraint precision, providing a 
$4.4\%$ estimate of $H_0$ based on 100 simulated systems, which is in agreement with the ground truth. This research highlights the potential of leveraging the capabilities of ML with glSNe systems to obtain a pipeline capable of fast and automated $H_0$ measurements.

\end{abstract}

\maketitle

\section{Introduction}\label{sec:introduction}

Recent measurements of the Hubble constant ($H_0$) from the Cosmic Microwave Background \citep[CMB; ][]{plank} radiation are in tension with observations from low-redshift probes, such as type Ia supernovae (SNe) calibrated by Cepheids \citep{shoes} or gravitationally lensed quasars \citep{h0licow}. The Hubble constant is a fundamental parameter for characterizing the cosmic expansion rate at any epoch, determining the age of the Universe, and calibrating the cosmic distance scale. Moreover, the persistent discrepancy between low-redshift measurements and those based on the CMB may hint at new physics beyond the standard $\Lambda$CDM model.
Nevertheless, this so called `Hubble tension' is lowered considerably by several other low-redshift measurements. Notably, the distance calibrations from the Tip of the Red Giant Branch (TRGB), as measured by the Carnegie-Chicago Hubble Project \citep{cornegie-chicago}, the re-analysis of seven gravitationally lensed quasars by the TDCOSMO collaboration \citep{tdcosmo} with less restrictive mass model priors than in \citet{h0licow} and a new approach of modeling extinction in SH0ES calibration galaxies \citep{2024MNRAS.533.2319W}. To determine whether new physics or residual systematics are behind the Hubble tension, there is a pressing need for new, independent measurements of $H_0$. 

Within the realm of late-time methods, gravitationally lensed SNe (glSNe) have emerged as a promising new probe for constraining $H_0$ \citep{Suyu2024}. By combining the relative time delays between multiple lensed images with the gravitational potential of the lensing galaxy, one can derive precise estimates of the Hubble constant \citep{Refsdal, treu2016cosmography, Birrer2024}. Compared to lensed quasars, the advantages of lensed SNe lie in their well-defined light curves and the possibility for follow-up observations to constrain the lens galaxy properties after the SN has faded away. 
While glSNe are very useful, they are exceptionally rare and challenging to detect. To date, only eight multiply-imaged lensed SNe have been found. Six of them have been discovered behind galaxy clusters: `SN Refsdal' \citep{sn_refsdal}, `SN Requiem' \citep{sn_requiem}, `AT2022riv' \citep{sn_kelly22}, `C22’ \citep{sn_chen22}, `SN H0pe' \citep{sn_hope}, and `SN Encore' \citep{sn_encore}, and two of them behind isolated elliptical galaxies: `iPTF16geu' \citep{sn_iptf} and `SN Zwicky' \citep{sn_zwicky}.
SN Refsdal and SN H0pe have provided Hubble constant measurements with uncertainties of $\sim 6\%$ \citep{Kelly2023_Refsdal_H0} and $\sim 9\%$ \citep{Pascale2024_H0pe_H0}, respectively, with the cluster mass models being the primary limitation on the precision. 
SN Encore is anticipated to achieve an $H_0$ measurement with approximately $\sim 10\%$ precision; however, the other glSNe either had insufficient data or too short time delays to provide meaningful constraints on the Hubble constant.

The study of lensed SNe is at a pivotal moment, as we will go from a handful current discoveries to an order-of-magnitude increase \citep{wojtak2019magnified, goldstein2019rates, oguri2010gravitationally, SainzdeMurieta2023, 44_sne} with the next generation of telescopes, such as the Vera C. Rubin Observatory \citep{ivezic2019lsst} and the Nancy Grace Roman Space Telescope \citep[e.g.,][]{spergel2015roman, pierel2020projected}. As these upcoming time-domain surveys will receive several terabytes worth of observations per night, processing the data becomes increasingly challenging. Machine learning (ML) methods have emerged as a way to make sense of the unprecedented volumes of data and to automate the detection and analysis of lensed sources. Specifically, ML models have been successfully employed to find strongly lensed systems in images from simulated and real transient surveys \citep{lanusse2018deeplens, schaefer2018deep, davies2019using, avestruz2019automated, cheng2020identifying,  huang2020finding, canameras2020holismokesII, huang2020discovering, gentile2021lenses, ramanah2021AIdriven}, to infer lens galaxy properties without costly Markov Chain Monte Carlo (MCMC) analyses \citep{hezaveh2017fast, wagnercarena2021hierarchical, park2021largescale, schuldt2020efficient}, and to obtain time-delay estimates from glSN light curves \citep{huber2021timedelay, hayes2024, Huber2024}. 
However, so far there has not been a complete, automated pipeline to take image time-series of glSNe as input and combine the inference of the time delays and lens model properties into an estimate of the Hubble constant. We present such a framework in this paper, in the context of observations from the Roman Space Telescope. We train an ensemble of five 3D convolutional neural networks (CNNs) on simulated image time-series and employ a simulation-based inference approach to quantify the uncertainties on the neural network predictions. By combining constraints from $100$ glSN systems, we obtain joint $H_0$ estimates, using \textit{only} Roman Space Telescope data in the $F087$-band and no follow-up observations, corresponding to a conservative scenario. We also show how an ensemble of neural networks can improve interpretability by providing insights into the error contribution from each parameter.

The remainder of this paper is organized as follows. 
Section~\ref{sec:theretical} introduces the theoretical framework of strong gravitational lensing. Section~\ref{sec:data} provides a description of our glSN image simulation procedure, assumptions regarding the lens mass profile, and SN light curves. 
Section~\ref{sec:ml} outlines the machine learning methodology, the model architecture, training, and validation processes. In Section~\ref{sec:results}, we present the results, evaluate the performance of the machine learning model in constraining $H_0$, and propose a method for combining measurements across multiple glSNe Ia.
Finally, we provide a discussion of the main findings of our study and possible avenues for future work in Section~\ref{Sec:Discussion}.

\section{Strong Gravitational Lensing}\label{sec:theretical}

Strong gravitational lensing occurs when the gravitational field of a massive object bends the light from a background source into multiple images. In this section, we describe the relevant quantities for the results in this work and the inference of the Hubble constant.
The Einstein radius, $\theta_E$, defines the angular scale of the lensing effect, indicating the typical separation between lensed images and marking the boundary within which a background source will be strongly lensed. It is given by:

\begin{equation}\label{eq:ein_rad}
\theta_{\textrm{E}} = \sqrt{\frac{4GM}{c^2} \frac{D_{\textrm{LS}}}{D_{\textrm{L}} D_{\textrm{S}} }},
\end{equation}
where $M$ is the mass of the lens, $G$ is the gravitational constant, $c$ is the speed of light, and $D_{\textrm{L}}$, $D_{\textrm{S}}$, and $D_{\textrm{LS}}$ are the angular diameter distances to the lens, the source, and between the lens and source, respectively.

Light rays from the background source take different paths to the lensed image positions, causing them to arrive at different times. The Fermat potential, $\tau$, describes the time delays experienced by light rays due to both the geometric path difference and the gravitational time dilation \citep[Shapiro delay;][]{shapiro1964GR}. The Fermat potential of an image at angular position $\boldsymbol{\theta}$ with corresponding source position $\boldsymbol{\beta}$ is given by:
\begin{equation} \label{eq:fermat_pot}
\tau(\bm{\theta}, \bm{\beta}) = \frac{1}{2} | \bm{\theta} - \bm{\beta} | ^2 - \psi(\bm{\theta}),
\end{equation}
where $\psi(\bm{\theta})$ is the projected 2D lens potential at the image position $\bm{\theta}$. The time delay between the multiple images, $\Delta t$, is directly related to the Fermat potential and the time-delay distance \citep{Refsdal, schneider1992lenses, suyu2010dissecting}, $D_{\Delta t}$, as:
\begin{equation} \label{eq:ddt}
\Delta t = \frac{D_{\Delta t}}{c} \Delta\tau(\bm{\theta}, \bm{\beta}),
\end{equation}
with $D_{\Delta t}$ defined as:
\begin{equation}
D_{\Delta t} = (1+z_{\textrm{lens}}) \frac{D_{\textrm{L}} D_{\textrm{S}} }{D_{\textrm{LS}} },
\label{eq:ddt_redshift}
\end{equation}
where $z_{\textrm{lens}}$ is the redshift of the lensing galaxy. The angular diameter distances $D_{\textrm{L}}$, $D_{\textrm{S}}$, and $D_{\textrm{LS}}$ are determined by integrals over the inverse Hubble parameter $H(z)$ as follows:
\begin{equation}
    D_{\textrm{A}} = \frac{c}{1+z}\int^{z}_{0}\frac{dz'}{H(z')}.
    \label{eq:dA}
\end{equation}
Assuming a flat $\Lambda$CDM cosmology, no significant contribution from radiation at the redshifts of interest and dark energy modeled as a cosmological constant, then $H(z)$ is given by:
\begin{equation}
    H(z) = H_0 \sqrt{\Omega_{m,0}(1+z^3) + \Omega_{\Lambda,0}},
    \label{eq:H}
\end{equation}
with $\Omega_{m,0}$ the present-day matter density parameter and $\Omega_{\Lambda,0}$ the present-day dark energy density parameter.
Thus, given that the redshifts of the source and lens are known, this relation implies that measurements of the time delays and the Fermat potential enable the determination of $H_0$. This method, known as time-delay cosmography, serves as a powerful technique for probing the expansion rate of the universe \citep{treu2016cosmography, Birrer2024}.

A significant challenge in time-delay cosmography is the mass-sheet degeneracy \citep[e.g.][]{Falco1985, Gorenstein1988, Saha2000, schneider2013MSD}, a phenomenon where different mass distributions change the time delays, but produce identical imaging observables.
This degeneracy arises when a uniform mass sheet with constant surface density is added to the lensing mass distribution, along with a scaling of the source plane coordinates. 
Mathematically, the mass-sheet degeneracy modifies the projected surface density, or convergence, $\kappa$, as:
\begin{equation}
\kappa \longrightarrow \lambda + (1-\lambda)\kappa,
\end{equation}
where $\lambda$ is a constant representing the additional mass sheet. The addition of this sheet affects the lensing potential and the inferred time delays, thus impacting the determination of $H_0$. GlSNe Ia possess several unique properties that help mitigate this degeneracy. Type Ia SNe are characterized by standardisable intrinsic luminosities, earning their designation as standard candles. For glSNe Ia, the difference between their intrinsic and observed luminosities directly constrains the lensing magnification, which helps to reduce the mass-sheet degeneracy \citep{kolatt1998magnification, oguri2003gravitational, birrer2021standardizable}. 
Although microlensing from stars in the lensing galaxy complicate this by adding stochastic uncertainty to the SNIa luminosities \citep{dobler2006microlensing, Yahalomi2017, foxley2018standardization, Weisenbach2021, Weisenbach_2024}, these effects can be incorporated in the lensing analysis to still obtain precise measurements of the lens galaxy mass slope \citep{Mortsell2020, Diego2022, Arendse2025}.
Furthermore, the transient nature of SNe facilitates follow-up observations of the lensing system after the SN has faded. As the SN dims, the lens galaxy and the lensed arcs of the host galaxy become more prominent, enabling detailed high-resolution imaging and spectroscopic follow-up observations, which further help to break the mass-sheet degeneracy \citep{ding2021improved, Birrer&Treu2021}. 
\section{Data Simulation and Generation}\label{sec:data}

Given the rarity and low number of glSNe detections to this date, glSN simulations are indispensable as they allow to train ML models, and enable the validation of analysis methods against known ground truth data.
To generate the dataset of glSNe Ia used in this study, we created realistic simulations, leveraging existing knowledge of supernova light curves, gravitational lensing, and the Roman Space Telescope characteristics. This section outlines the methodology for simulating the light curves, lensing effects, and observational conditions, culminating in a dataset of 50,000 image time-series of glSNe Ia systems. The code developed and employed for the simulations is publicly available on a  \href{https://github.com/G-GoncaIves/thesis}{GitHub repository}.

\subsection{Supernova Light Curve Modeling}\label{subsec:lightcurve}

The light curves of SNe Ia are simulated using the Spectral Adaptive Lightcurve Template 3 \citep[\texttt{SALT3};][]{guy2007salt2, salt3}, which uses a combination of principal components to describe the flux evolution of SNe Ia. These components are derived from a training dataset of observed SNe Ia spectra and light curves.

These components capture the mean behaviour and primary variations in supernova light curves and spectra. This approach enables SALT3 to accurately standardize the light curves by incorporating detailed spectral information and accounting for variability. The main parameters that contribute to the SNe Ia variability are the stretch, $x_1$, which relates to the width of the light curve, and the color parameter, $c$, which is a combination of the intrinsic SN color and reddening by dust in the host galaxy. They affect the absolute peak brightness of the SN, such that broader and bluer SNe Ia are brighter. This effect is quantified in the Tripp formula \citep{tripp}, which expresses the distance modulus DM as:

\begin{equation} 
\textrm{DM} = m + \alpha x_1 - \beta c - \mathcal{N}(M_0, 0.12), 
\end{equation}

where $m$ is the apparent peak magnitude, $\alpha$ and $\beta$ are coefficients quantifying the dependence of absolute magnitude on stretch and color, respectively, and $M_0$ is the fiducial absolute magnitude for a supernova with $x_1=0$ and $c=0$. We assume $M_0 = -19.4$ in the $B$-band, corresponding to a Hubble constant of 67.7 $\mathrm{~km~s^{-1}~Mpc^{-1}}$

For our simulations, we sample $x_1$ and $c$ from asymmetric Gaussian distributions determined in \cite{xc_dist}, which are provided in Table~\ref{tab:sampling_distributions}. We assume $\alpha = 0.14$ and $\beta = 3.1$ \citep{xc_dist}. 
We use the SALT3 template implemented in the open source software \texttt{SNcosmo} \citep{sncosmo} to simulate the SNe Ia light curves.

\begin{table}[h!]
\centering
\begin{tabular}{ll}
\hline
\\
\textbf{Parameter} & \textbf{Distribution} \\
\\
\hline
\\
\textbf{Light curve} \\
Stretch & $x_1 \sim \mathcal{N^A}~$$(~0.945,~1.553,~0.257~)$ \\
Color   & $c \sim \mathcal{N^A}~$$(~-0.043,~0.052,~0.107~)$ \\
Absolute magnitude & $M_0 \sim \mathcal{N}~$$(~-19.4,~0.12~)$ \\
Stretch coefficient & $ \alpha \sim \delta~(~x-0.14~)$ \\
Color coefficient & $ \beta \sim \delta~(~x - 3.1~)$ \\
\\
\textbf{Lensing galaxy} \\
Axis ratio & $ q_{\mathrm{lens}} \sim \mathcal{N}~(~0.7,~0.15)$ \\
Orientation angle [rad] & $ \mathrm{\phi_{lens}} \sim \mathcal{U}~(~-\frac{\pi}{2},~\frac{\pi}{2})$ \\
\\
\textbf{Geometrical configuration} \\
Source $x$-position [arcsec] & $ x_\mathrm{source} \sim \mathcal{U}~(~-1,~1)$ \\
Source $y$-position [arcsec] & $ y_\mathrm{source} \sim \mathcal{U}~(~-1,~1)$ \\
\\
\hline \\
\textbf{Simulated telescope properties} \\ 
\\ \hline \\
PSF type & Gaussian \\
PSF's FWHM~[arcsec] & 0.073\\ 
Cadence~[days] & 5 \\
Magnitude Zero Point & 26.30 \\
Read Noise~[$e^{-}$] & 15.5 \\
Sky Brightness [$\textrm{arcsec}^{-2}$] & 22.93 \\
\\ 
\hline
\end{tabular}
\caption{Sampling distributions for the parameters and telescope properties (corresponding to the Roman Space Telescope) used in the simulations. $\mathcal{N}^{A}(\mu, \sigma_{+}, \sigma_{-})$ denotes an Asymmetric Gaussian distribution, where $\sigma_+$ and $\sigma_{-}$ represent the standard deviation to the right and left of the mean $\mu$, respectively. $\mathcal{N}(\mu,\sigma)$ denotes a Gaussian distribution with mean $\mu$ and standard deviation $\sigma$. $\mathcal{U}$ denotes a Uniform distribution and $\delta(x-x_0)$ denotes a Dirac delta distribution. The Einstein radius, $\theta_{\textrm{E}}$, source redshift, $z_{\textrm{source}}$, and lens redshift, $z_{\textrm{lens}}$, are not included in this table because they are sampled from a joint distribution from \citet{wojtak2019magnified}.}
\label{tab:sampling_distributions}
\end{table}

\subsection{Lens galaxy mass profile}

We model the mass distribution of the lens galaxy as a Singular Isothermal Ellipsoid \citep[SIE;][]{sie}. The SIE has been shown to be a reasonable approximation for massive elliptical galaxies \citep{Auger2009}, which is sufficient for our first application of this ML pipeline.
The convergence of the SIE is given by:
\begin{equation}
    \kappa (\theta_1, \theta_2) = \frac{\theta_{\textrm{E}} }{2\sqrt{\theta_1^2 + q_{\rm{lens}}^2 \theta_2^2}},
    \label{eq:sie}
\end{equation}
where the terms $\theta_1$ and $\theta_2$ are the angular coordinates on the sky, specifying the positions in the plane of the lens where the convergence is being evaluated. The coordinates are centered on the position of the lens center, and rotated by the lens orientation angle $\phi_\textrm{lens}$.
$\theta_{\textrm{E}}$ is the Einstein radius (as defined in Eq.~\ref{eq:ein_rad}) and $q_{\rm{lens}}$ is the ratio of the minor axis to the major axis of the elliptical mass distribution of the lens. This parameter describes the ellipticity of the lensing potential, where a value of $q_{\rm{lens}}=1$ corresponds to a circularly symmetric (spherical) lens, and values less than 1 indicate increasing ellipticity. 
The axis ratio, $q_{\rm{lens}}$, together with the system's orientation angle, $\phi_\textrm{lens}$ can be converted into:
\begin{equation}
\begin{aligned}
    e_1 = \frac{1-q_{\rm{lens}}}{1+q_{\rm{lens}}} \cos(2 \phi_\textrm{lens}) \\
    e_2 = \frac{1-q_{\rm{lens}}}{1+q_{\rm{lens}}}  \sin(2 \phi_\textrm{lens}),
    \label{eq:e_to_q}
\end{aligned}
\end{equation}
where $e_1$ and $e_2$ represent the ellipticity in a Cartesian coordinate system. This is often more practical than using $q_{\rm{lens}}$ and $\phi_\textrm{lens}$ directly, especially to circumvent issues arising from cyclic boundary conditions due to the 2$\pi$-periodic property of the angles when we train our network. 

We sample most parameters independently from the distributions listed in Table~\ref{tab:sampling_distributions}. 
For the Einstein radius, the lens redshift and the supernova redshift, we use a 3D probability distribution obtained by applying a kernel density estimator to a Monte Carlo sample of detectable glSNe computed by \citet{wojtak2019magnified}. The original simulation employs observationally motivated type Ia supernova rate as a function of redshift and the mass function of lens galaxies, which are the two primary factors relevant for predicting lensing configurations in terms of the Einstein radius and redshifts of the supernova and lens galaxy.

We use the open-source lens modeling software \texttt{Lenstronomy}\footnote{\href{https://lenstronomy.readthedocs.io/en/latest/}{Lenstronomy}} \citep{lenstronomy, Birrer2021LenstronomyII} to solve the lens equation, reconstruct observables such as image positions, time delays, and magnifications, and to create the final image-level observations. For the variability of the glSN images, we generate synthetic light curves with \texttt{SNCosmo} usig the \texttt{SALT3} template, as described in Section~\ref{subsec:lightcurve}. In all our simulations, we assume a standard flat $\Lambda$CDM model with $H_0 = 67.7~\textrm{km}~\textrm{s}^{-1}~\textrm{Mpc}^{-1}$, $\mathrm{\Omega_{m,0}=0.3}$ and $\mathrm{\Omega_{\Lambda,0}=0.7}$.

\subsection{Nancy Grace Roman Space Telescope}
\label{subsect:Roman_telescope}

The Nancy Grace Roman Space Telescope (hereafter the Roman Space Telescope), to be launched in 2026, combines wide-field imaging and spectroscopy to map billions of galaxies, probe dark energy, and discover thousands of exoplanets. Its advanced features, including a large field of view (0.28 $\mathrm{deg^2}$), high angular resolution (0.1 arcsecs), and stable space-based observations are ideal for detecting glSNe Ia and studying them in detail. \cite{sne_detection} predicts that the Roman Space Telescope will discover 11 glSNe Ia during its mission. However, this number could increase significantly if the telescope is used to follow up on glSNe Ia, which was discovered by the Vera Rubin Observatory's Legacy Survey of Space and Time (LSST). Forecasts based on the full LSST observing strategy estimate that LSST will detect approximately 44 glSNe Ia annually \citep{44_sne}. By using the Roman Space Telescope to follow up these discoveries, we can build a substantial dataset of high-quality glSN observations. In this work, we simulate such a dataset.

For the first proof of concept of our method, we limit ourselves to Roman's $F087$ band, which spans the wavelength range that includes much of the light from high-redshift SNIa. Observational effects, such as the point spread function (PSF), were modeled using the $F087$ band's Gaussian PSF full width at half maximum (FWHM) of 0.073 arcsec\footnote{\href{https://roman.gsfc.nasa.gov/science/WFI_technical.html}{Roman Performance Information}}. A five-day observational cadence was assumed based on preliminary mission specifications, for the High Latitude Time Domain Survey\footnote{\href{https://www.stsci.edu/files/live/sites/www/files/home/roman/_documents/roman-science-and-technical-overview.pdf}{Roman Science and Technical Overview}}.

\subsection{Dataset Generation}

\begin{figure}[H]
    \centering
    \includegraphics[width=\linewidth]{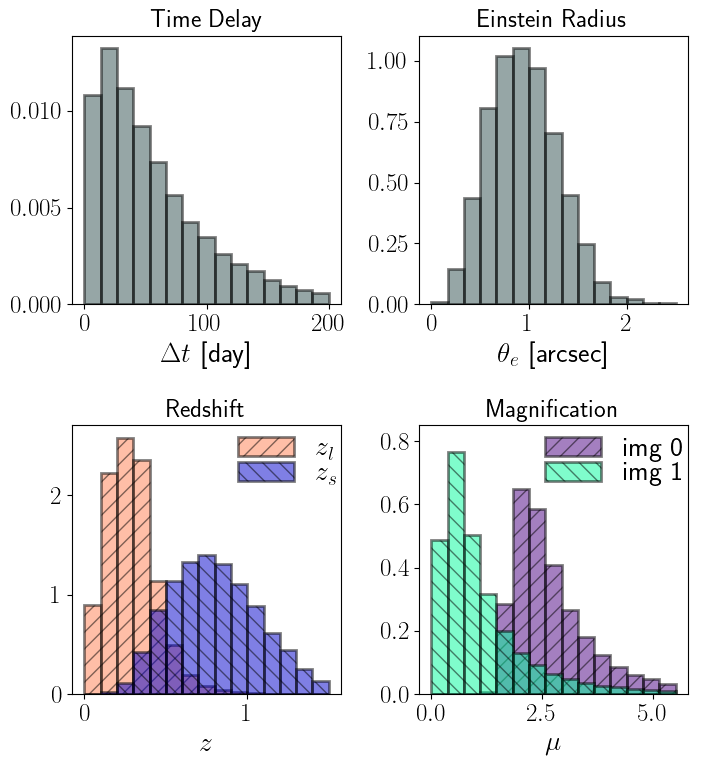}
    \caption{\small Distributions of the properties of the simulated glSNe Ia, resulting from the simulated doubly imaged dataset. \\}
    \label{fig:gen_dist_im}
\end{figure}

The combined models of the supernovae, lensing systems, and observational effects were used to generate a dataset of 50,000 simulated glSNe Ia systems.  
Each simulated system, once completed, was stored as an individual element of our dataset, with the properties of its resultant images recorded and summarized in Figure~\ref{fig:gen_dist_im}.

\begin{figure}
    \centering
    \begin{subfigure}{\linewidth}
        \centering
        \includegraphics[width=\linewidth]{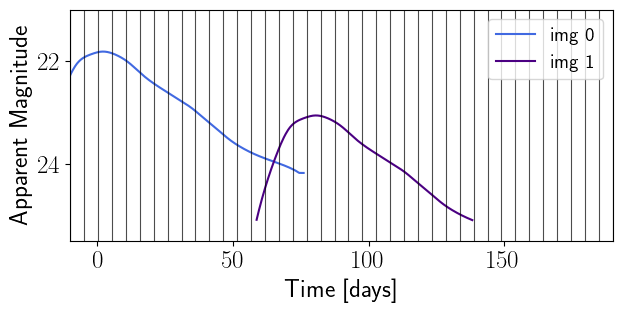}
        \caption{ }
    \end{subfigure}
    \hfill
    \begin{subfigure}{\linewidth}
        \centering
        \includegraphics[width=\linewidth]{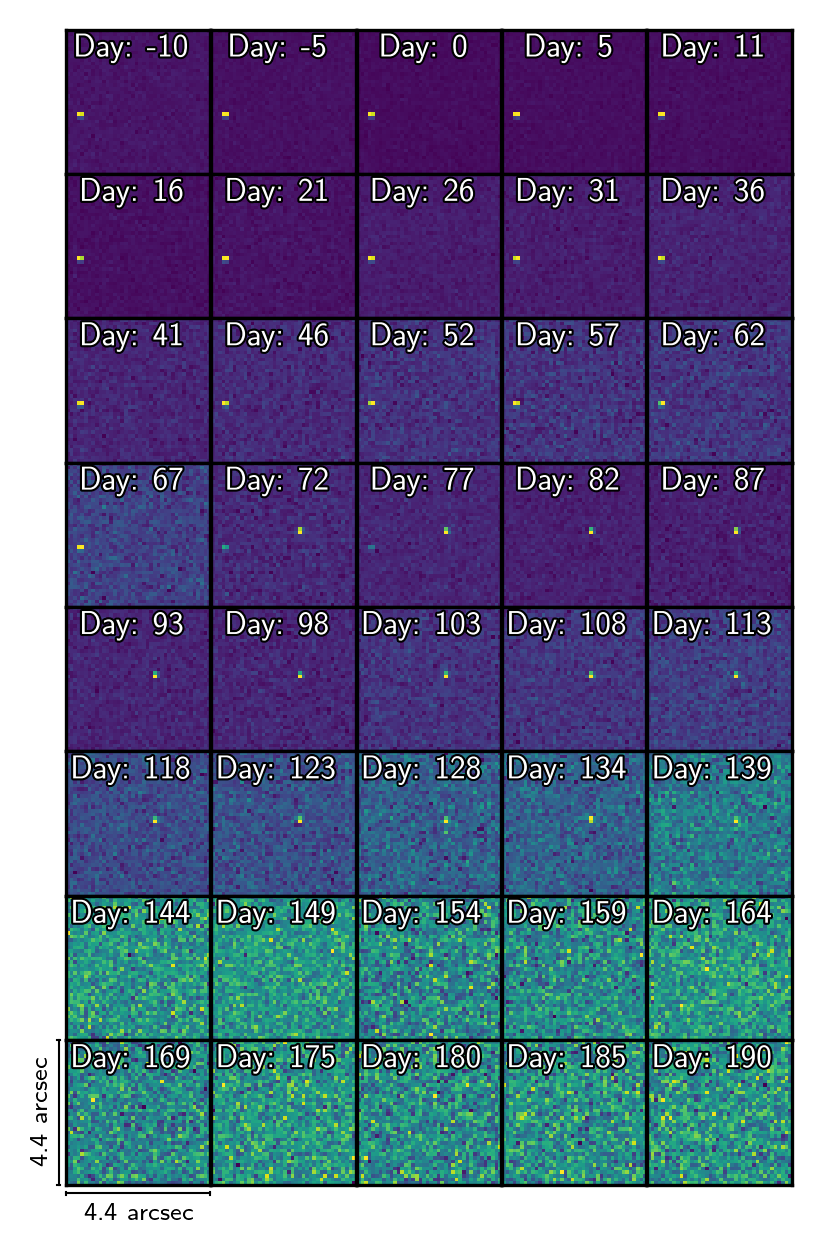}
        \caption{ }
    \end{subfigure}
    \caption{\small Example of an element in our doubly imaged dataset. Here a simulated glSN Ia system is shown, where: \textbf{(a)} the images' light curves are shown, with vertical grey lines indicating the time at which the images were simulated; \textbf{(b)} the time-series (i.e. the element), comprised of the images simulated at all the aforementioned times, is shown. Each element holds the evolution of a glSN, where observations are taken every five days. The glSN Ia system shown here has: $\theta_{\textrm{E}}$ = 1.157, $z_l$ = 0.232, $z_s$ = 0.527, $\Delta t$ = 78.48 days, $\mu_{\textrm{img}~0}$ = 2.688 and $\mu_{\textrm{img}~1}$ = 0.854.}
    \label{fig:sample}
\end{figure}

Each element of the dataset represents a sequence of observations of the same gravitationally lensed supernova system, thereby capturing its temporal evolution. Specifically, each element consists of an image time-series corresponding to the telescope's perspective at discrete time intervals. In this study, we modeled an observational period spanning 200 days, employing a telescope cadence of five days, thus generating time-series datasets consisting of 40 sequential images for every glSNe Ia event. Each image possesses an angular dimension of 4.4 by 4.4 arcseconds. Figure~\ref{fig:sample} provides an example of a dataset element, including the corresponding light curve.

The majority of lensed supernovae found by the Roman Space Telescope are expected to be doubles; systems with two visible images. They are also the ones with typically higher time delays, and therefore they are well-suited for cosmology. For those reasons, we only simulate doubles and leave the exploration of using quads with our pipeline for future work.

The simulated data employs difference imaging, a standard technique for identifying transient phenomena such as supernovae. This method involves subtracting a reference image from subsequent observations, thereby isolating transient sources like the supernova while removing the contributions of the lensing and host galaxies. In this work, a perfect subtraction of the lens and host galaxy light was assumed. The next step would be to simulate a glSN dataset using a realistic difference imaging pipeline.

The high spatial resolution of the telescope, characterized by a small point spread function (PSF), ensures that the glSNe appear as distinct point sources. Each SN image occupies only a few pixels, facilitating precise measurement and analysis of the system's properties.

\section{Machine learning model}\label{sec:ml}

In our work, we employ an ensemble of five 3D convolutional neural networks, as described in more detail in Section~\ref{subsect:CNNs}, to process the image time-series, in combination with a simulation-based inference (SBI) framework, as outlined in Section~\ref{subsect:SBI}, to infer robust uncertainties on the network predictions. Section~\ref{subsect:H0_inference_pipeline} describes the various steps of our complete $H_0$ inference pipeline, while Section~\ref{subsect:architecture_training} outlines the training procedure.

\subsection{Convolutional neural networks}
\label{subsect:CNNs}

A neural network is a trainable approximation of a model that performs some form of data compression (or dimensionality reduction) to extract meaningful, informative features from a given input dataset. In essence, it maps some input data $\myvec{D}$ to a prediction $\tilde{\myvec{d}}$ of a desired target $\myvec{d}$, i.e. $\mathbb{NN} (\myvec{\omega}, \myvec{\eta}): \myvec{D} \rightarrow \tilde{\myvec{d}}$, where $\myvec{\omega}$ and $\myvec{\eta}$ correspond to the weights of the network and hyperparameters, respectively, as explained below.

Convolutional neural networks \citep[CNNs,][]{lecun1995convolutional, lecun1998gradient} are a class of neural networks that are particularly well-suited for the analysis of visual data. They are designed to extract spatially-correlated information from data by sliding a \textit{kernel} across the image to check if a feature is present, an operation that is known as a \textit{convolution}. This process yields a \textit{feature map} which contains high values in the pixels that match the pattern encoded in the kernel. In the case of a 3D CNN, as in our work, the input data, kernels, and feature maps are 3D matrices. By stacking 2D images to form a 3D representation, we incorporate the temporal dimension in our input data, such that the 3D CNN is also sensitive to temporal correlations in the data.  In general, a standard CNN contains an ensemble of the following types of layers:
\begin{itemize}
\setlength\itemsep{0.01em}
\item \textbf{Convolutional layer}. A convolutional layer usually employs multiple kernels to extract several features from the input image, resulting in a set of feature maps which are then fed as inputs to the next layer.
\item \textbf{Activation function}. In order to incorporate some non-linearity in the network and to restrict the output values to a certain range, an activation function is used after each convolutional layer. The network used in this work employs a \texttt{tanh} (hyperbolic tangent) activation function.
\item \textbf{Pooling}. The function of a pooling layer is to downsample the feature maps, often by taking the maximum or average of the window defined by the kernel \citep[the \textit{receptive field,}][]{goodfellow2016deep}. This process renders the resulting downsampled feature maps more robust to overfitting and to changes in the position of a feature in the image. 
\item \textbf{Linear layer}. After the feature maps are flattened to a 1D array, they are fed into a set of linear layers to ultimately generate an output \citep{lecun2015deep}.
\end{itemize}

The more convolutional layers are stacked together, the better the network can capture intricate features and complex patterns in the input data. As we go deeper into the network, the size of the receptive field becomes larger and the layers become sensitive to features on increasingly larger scales.

The complexity of the network depends on its weights $\myvec{\omega}$ and set of hyperparameters $\myvec{\eta}$, the latter referring to the network architecture, initialization of the weights, and type of activation and loss functions.
During the training phase of the network, the weights and hyperparameters are optimized according to stochastic gradient descent to minimize a loss function, which is often a negative log-likelihood, $- \ln \mathcal{L} (\tilde{\myvec{d}} | \myvec{D}, {\myvec{\omega}}, {\myvec{\eta}})$. The loss function quantifies how close the prediction $\tilde{\myvec{d}}$ is to the desired target ${\myvec{d}}$. As such, the training process is equivalent to determining a maximum likelihood estimate of the weights, resulting purely in a single point estimate by the neural network, which would ideally correspond to the global minimum on the likelihood surface. However, in practice, this is generally not the case, as the likelihood
surface is typically extremely complex, degenerate, and non-convex. Therefore, it is much more likely that the weights will only converge to a local minimum on the likelihood surface, with the subsequent local minimum dictated to some extent by the initialization of the weights. This is one of the major drawbacks of using neural networks because it means that the network point estimates never truly correspond to the desired target ${\myvec{d}}$. Simulation-based inference, as described in the following subsection, provides a way to mitigate this crucial limitation and to obtain scientifically accurate predictions, including reliable uncertainties, of the true targets $\myvec{d}$ given the input data $\myvec{D}$.

\subsection{Simulation-based inference}
\label{subsect:SBI}

While simulation-based inference \citep[][hereafter SBI]{cranmer2019frontier} comprises a panoply of statistical inference methods, the underlying theme relies on a simulator whose task is to generate high-fidelity simulations that can eventually be compared with observations. The key aspect of SBI is that it allows the probability density for a given observation to be estimated directly from the distribution of summary statistics of the simulated realizations. A neural network is ideally suited for extracting informative summaries of the data, while a standard density estimator, such as a Gaussian kernel density estimator (KDE), or a neural density estimator \citep{germain2015made, papamakarios2017MAF}, can be used to approximate the desired likelihood or posterior from the distribution of network predicted summaries.

\begin{figure}
	\centering
		{\includegraphics[width=\hsize,clip=true]{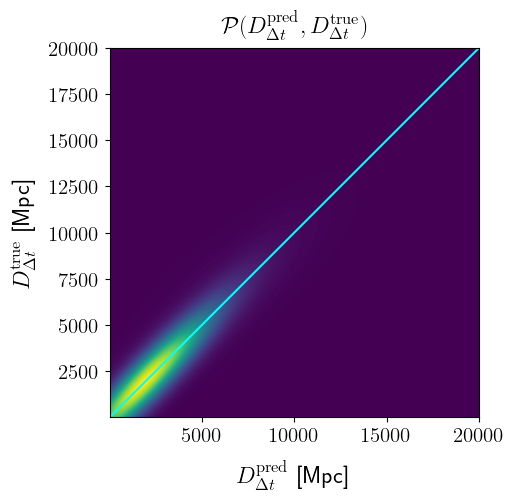}}
	\caption{\small Joint 2D PDF of predicted and true $D_{\Delta t}$ values, computed using a set of 24,245 doubly imaged supernovae. We obtain the approximate posterior $\mathcal{P}(D_{\Delta t}^{\textrm{true}} | D_{\Delta t}^{\textrm{pred}})$ for a particular lens system by making a vertical slice at the predicted value of $D_{\Delta t}$. \\}
	\label{fig:true_pred_kde}
\end{figure}

For our setup, we characterize the joint probability density function (PDF) of the predicted and ground truth values of the time-delay distance: $\mathcal{P}(D_{\Delta t}^{\textrm{pred}}, D_{\Delta t}^{\textrm{true}})$, using a Gaussian KDE. The 2D PDF for a set of 24245 doubly imaged supernova systems is depicted in Figure~\ref{fig:true_pred_kde}. Finally, for any observed image time-series $\myvec{D}_{\mathrm{obs}}$, we obtain the approximate posterior by slicing the joint PDF at the predicted $D_{\Delta t}$ value corresponding to the network predicted summaries, $\mathbb{NN}({\myvec{\omega}}, {\myvec{\eta}}): \myvec{D}_{\mathrm{obs}} \rightarrow \tilde{\myvec{d}}_{\mathrm{obs}}$, as follows:
\begin{equation}
	 \mathcal{P}(D_{\Delta t} | \tilde{\myvec{d}}_{\mathrm{obs}}) \approx \mathcal{P}( D_{\Delta t} | \myvec{D}_{\mathrm{obs}}, {\myvec{\omega}}, {\myvec{\eta}} ) .
	\label{eq:SBI_posterior}
\end{equation}

\begin{figure*}
    \centering
    \includegraphics[width=\linewidth]{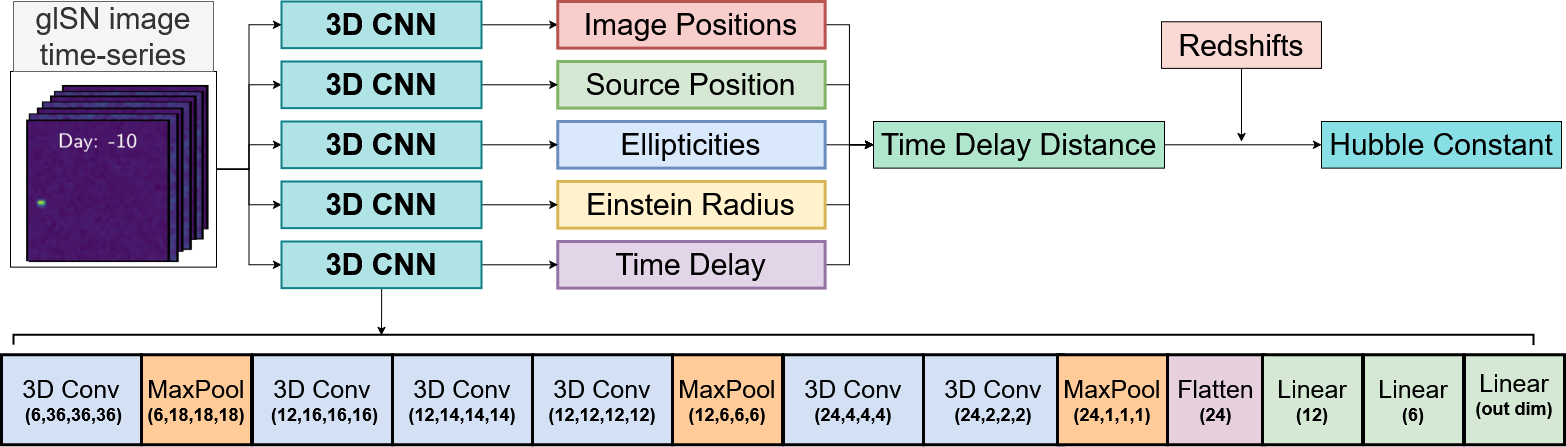}
    \caption{Illustration of the ensemble architecture used in this study. The ensemble processes a sequence of images through five independent 3D CNNs, each designed to infer a specific parameter: image positions, source position, ellipticities, Einstein radius, and time delay. Each layer in the CNNs is followed by a \texttt{tanh} activation function, ensuring bounded outputs suitable for parameter estimation. The shared architecture of the CNNs is detailed at the bottom of the figure, with the primary distinction being the dimensionality of the final layer, which is tailored to each output parameter. These individual outputs are subsequently utilized to compute the time-delay distance, which is combined with redshift measurements to derive an estimate of the Hubble constant.}
    \label{fig:ensemble}
\end{figure*}

The primary advantage of our implementation of the SBI framework is that it yields a posterior of the parameter of interest with statistically consistent uncertainties, despite the posterior being conditional on the trained neural network weights ${\myvec{\omega}}$ and chosen hyperparameters ${\myvec{\eta}}$. This framework ensures that the uncertainties in the network predictions are not underestimated as a sub-optimal network will only result in inflated uncertainties. Furthermore, only one application of the density estimator is required to compute the joint PDF of network summaries and parameters. The desired approximate posterior for any given observation can then be obtained almost instantly by slicing this joint PDF at the corresponding network-predicted summary. Although there is a choice of density estimator that in turn depends on some chosen hyperparameters, the Gaussian KDE is a reasonably robust option as it is not highly sensitive to the hyperparameter settings. The other well-known caveat of SBI approaches is that a large number of simulations are required to adequately characterize the joint PDF, and this can be mitigated with a sufficiently large dataset.

\subsection{$\mathbf{\textit{H}_0}$ inference pipeline}
\label{subsect:H0_inference_pipeline}

For doubly-imaged glSNe, the time-delay distance $D_{\Delta t}$ can be determined analytically from estimates of the lens model ($\theta_{\textrm{E}}, \ e_1, \ e_2$), the image positions ($\myvec{x}_{\textrm{im}}, \ \myvec{y}_{\textrm{im}}$), the source position (${x}_{\textrm{src}}, \ {y}_{\textrm{src}}$) and the time delay ($\Delta t$) between the images, via Eq.~\ref{eq:ddt}. 
In this work, an ensemble of five CNNs with the same architecture, as shown in Fig.~\ref{fig:ensemble}, is used to obtain predictions for $\tilde{\myvec{d}} \equiv \left\{ \Delta t_{\textrm{max}}, \theta_{\mathrm{E}}, e_1, e_2, x_{\textrm{src}}, y_{\textrm{src}}, \myvec{x}_{\textrm{im}}, \myvec{y}_{\textrm{im}} \right\}$, which are subsequently converted analytically into an estimate of the time-delay distance.   
Ensemble models have increased interpretability, since each model's inputs and outputs can be probed. It then becomes possible to investigate the ensemble and find how each model's performance contributed to the end result. 

The final step of the pipeline is to convert $D_{\Delta t}$ analytically into $H_0$ via Eqs.~\ref{eq:ddt_redshift}, \ref{eq:dA}, \ref{eq:H} and a cosmological model to relate angular diameter distances to the Hubble constant. We adopt a standard $\Lambda$CDM model with $\Omega_{\textrm{m},0} = 0.3$, $\Omega_{\Lambda,0} = 0.7$ and $\Omega_{\textrm{k}} = 0.0$, which has also been employed in the image simulation procedure.
Additionally, this step requires information about the source and the lens redshifts, for which we supply the ground truth values of $z_{\textrm{lens}}$ and $z_{\textrm{src}}$. We thereby assume that spectroscopic redshift measurements of the lens and the host galaxy will be carried out after the supernova has faded away. The redshifts are only provided in this final step and not fed as input to the CNNs, ensuring no leakage of cosmological information into the final results. 

In essence, the network predictions ${\tilde{\myvec{d}}}$ are converted analytically into estimates of $D_{\Delta t}$ and $H_0$, after which full posteriors are obtained through our SBI framework. Fig.~\ref{fig:ensemble} illustrates a schematic overview of our pipeline to infer a single $H_0$ estimate from the image time-series of lensed supernova.

\subsection{Training}
\label{subsect:architecture_training}

After some initial testing, it was found that using 25,000 of the simulated glSNe Ia systems for training\footnote{Out of the 25,000, 15,000 were used for the first phase, where the model's weights are updated via back-propagation, and the remaining 10,000 were used to validate the evolution of the model at each epoch.} was the right trade-off between achieved performance and computational cost incurred. Additionally, 5,000 were used to test all models and produce the results in the upcoming subsections. The training of each ensemble model was done separately, but on exactly the same data and under the same conditions. Since the goal is to train regression models, the training was done such that it minimizes the Mean Squared Error (MSE), which is defined as:

\begin{equation}
    \textrm{MSE} = \frac{1}{N_{D}}\sum_{i}^{N_{D}}~(~y_i~-~\hat{y}_i~)^2,
    \label{eq:MSE}
\end{equation}
where $N_D$ is the number of simulated glSNe Ia systems used, $y_i$ is the prediction for a given system and $\hat{y}_i$ the corresponding ground truth or target. 
\section{Results}\label{sec:results}

In this section, the performance of the 3D CNN ensemble model is evaluated by first examining the individual behavior of each model within the ensemble, and then assessing the ensemble's overall performance. We present our final results as a joint estimate of the Hubble constant from a set of 100 glSNe Ia from our test set.

\begin{figure*}
    \centering
    \includegraphics[width=\linewidth]{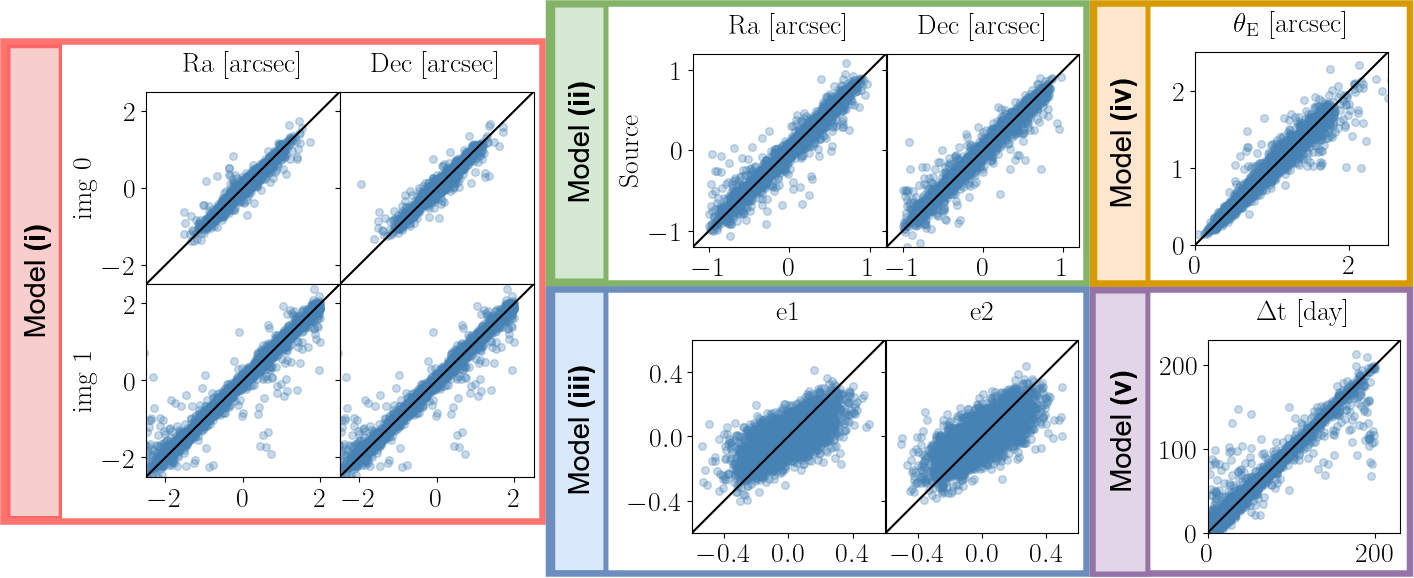}
    \caption{Comparison between predicted (y-axis) versus ground truth (x-axis) for all parameters predicted by the 3D CNN ensemble model. The colored boxes indicate parameters predicted by the same model and they are as follows: \textbf{(i)} Inside the red square are the predictions for all the image position coordinates; \textbf{(ii)} Inside the green square are the source position coordinates;  \textbf{(iii)} Inside the blue square are the Ellipticities predictions, $e_1$ and $e$; \textbf{(iv)} Inside the yellow square are the predictions for the Einstein Radius, $\theta_E$;  and, finally, \textbf{(v)} Inside the purple square are the time delay, $\Delta t$, predictions.}
    \label{fig:model_preds}
\end{figure*}

Rather than presenting the MSE  (eq.~\ref{eq:MSE}) for all models, the Mean Error (ME) is provided to offer a more intuitive metric, defined as:

\begin{equation}
    \textrm{ME} = \frac{1}{N_{D}}~\sum_{i}^{N_{D}}~(~y_i~-~\hat{y}_i~).
\end{equation}

Since the training procedure minimizes the MSE, the ME will also be minimized. Thus, this metric is still faithful to the minimized quantity but is more intuitive.

\subsection{Models Performance}

The performance of each model within the ensemble was evaluated regarding the estimation of the respective parameters, as illustrated in Fig.~\ref{fig:model_preds}. Analyzing the results, the following observations can be made for each model:

\begin{itemize}
    \item \textbf{Source Positions:} The model estimated the supernova's position within its host galaxy with an ME of 0.045 arcseconds. This accuracy underscores the model's proficiency in effectively determining the coordinates of the source.
    \item \textbf{Image Positions:} For the positions of the lensed images, the model achieved an ME of 0.071 arcsececonds. The predictions demonstrate some limitations due to challenges in capturing spatial features near the boundaries of the input data.
    \item \textbf{Ellipticities:} The model achieved an ME of 0.075 in predicting ellipticity components \((e1\) and \(e2)\). Inferring ellipticities from our simulated dataset, which only contains difference images of doubly-imaged glSNe Ia 
    and excludes light contributions from the lens and host galaxies, poses a challenge for the model.
    Despite these difficulties, the model effectively learned the overarching trends in ellipticity.
    \item \textbf{Einstein Radius:} For the Einstein radius ($\theta_E$), the model demonstrated an ME of 0.054 arcsec. Errors were found to be correlated with the absolute value of $\theta_E$, which can be understood since systems with larger $\theta_E$ have features close to the image boundaries, where CNNs are known to struggle with extracting reliable information.
    \item \textbf{Time Delay}: The model trained to predict the time delay between images ($\Delta t$) exhibited an ME of 4.08 days. 
    The ME aligns closely with the simulation cadence of five days, suggesting that the model's performance is influenced by this temporal resolution. 
    As expected, small time delays were challenging to estimate accurately. On the other hand, very large time delays often led to increased errors as well, which can be attributed to the 200-day limit in the simulation data. As a consequence of that, glSNe Ia with time delays approaching 200 days fall outside the simulated range. However, as can be seen in Fig.~\ref{fig:gen_dist_im}, such cases represent only a small fraction of the total dataset.
\end{itemize}

\subsection{Estimating the time-delay distance}\label{subsec:point_estimates}

\begin{figure}
    \centering
    \includegraphics[width=\linewidth]{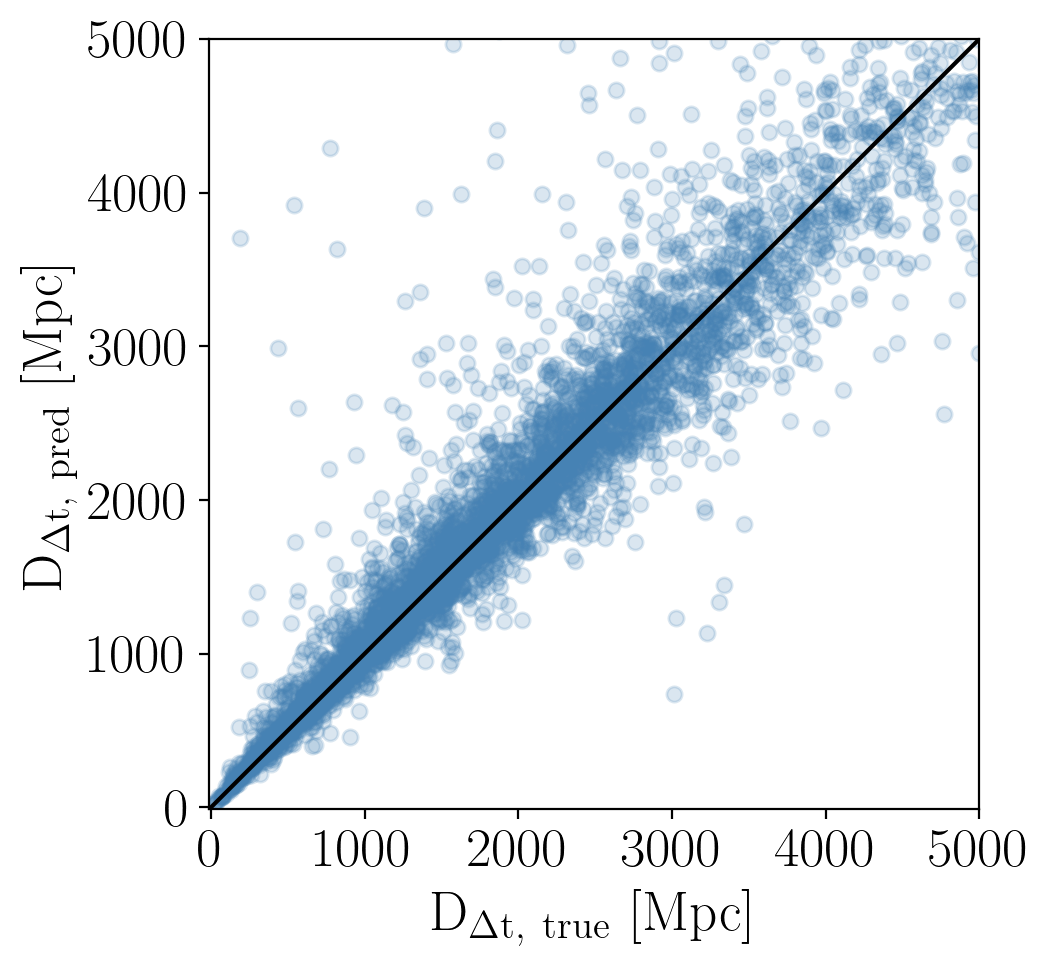}
    \caption{\small Estimates for $D_{\Delta t}$ calculated for all glSNe Ia versus their corresponding ground truths. \\}
    \label{fig:tdd_estimates}
\end{figure}

Having trained the ensemble model to accurately predict the necessary parameters, it is now possible to get point estimates of the time delay distance, $D_{\Delta t}$, for all simulated glSNe Ia systems in the dataset. The first step is to use Eq.~\ref{eq:fermat_pot} to calculate the Fermat potential, $\tau$, which, together with the model's estimate of the time delay, $\Delta t$, can be used to calculate $D_{\Delta t}$ via Eq.~\ref{eq:ddt}. Following this procedure, the $D_{\Delta t}$ estimates for all systems are obtained and shown in Fig.~\ref{fig:tdd_estimates}.

\subsection{Estimating the Hubble constant}

\begin{figure}[h!]
    \centering
    \includegraphics[width=\linewidth]{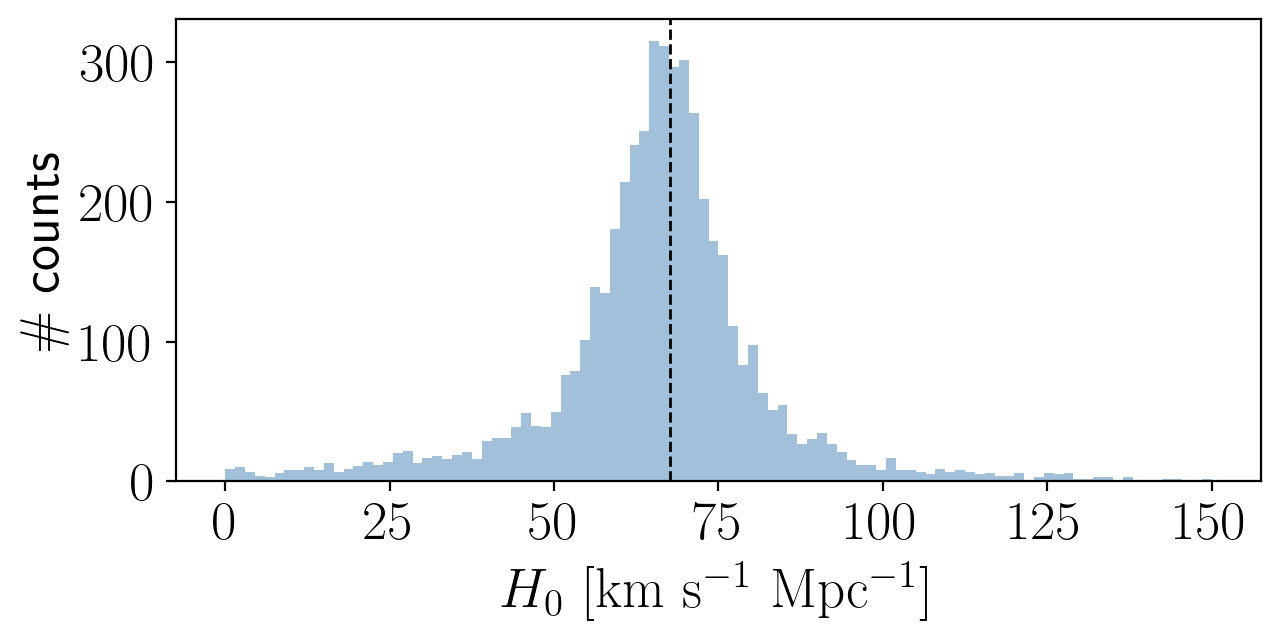}
    \caption{\small Histogram of $H_0$ point estimates predicted by the network for each glSN system  The vertical black dashed line indicates the simulation's ground truth value, $H_0 = 67.7~\mathrm{km~s^{-1}~Mpc^{-1}}$. The distribution has a median of $66.46\mathrm{~km~s^{-1}~Mpc^{-1}}$ and a SD of $17.53\mathrm{~km~s^{-1}~Mpc^{-1}}$. \\}
    \label{fig:h0_hist}
\end{figure}

To determine the Hubble constant, $H_0$, from time-delay distance estimates, the redshifts of both the lens galaxy and the supernova host galaxy must be known. The Roman Space Telescope will provide high-resolution photometric imaging and slitless spectroscopy, but its spectroscopic sensitivity begins at wavelengths $\gtrsim 0.75~\mu m$. As a result, key spectral features of Type Ia supernovae, typically found in the rest-frame optical ($0.33–0.5~\mu m$), will only fall within Roman’s spectroscopic bandpass for supernovae at redshifts $z \gtrsim 1$. For systems at lower redshift or for events that are too faint, follow-up observations with other observatories will be necessary to obtain accurate redshifts, such as the Hubble Space Telescope or the James Webb Space Telescope. $H_0$ is then calculated by using the $D_{\Delta t}$ point estimates and Eqs.~\ref{eq:ddt_redshift}, \ref{eq:dA}, and \ref{eq:H}, where $H_0$ is a free parameter and $\Omega_{m,0}$ and $\Omega_{\Lambda,0}$ are set, according to \cite{plank_h0}, to 0.3 and 0.7, respectively. The resulting estimates for the Hubble constant are shown in Fig.~\ref{fig:h0_hist}. The estimates peak around the $H_0$ value used as ground truth for the simulations ($H_0 = 67.7\mathrm{~km~s^{-1}~Mpc^{-1}}$) with a median of $H_0 = 66.46\mathrm{~km~s^{-1}~Mpc^{-1}}$. However, the distribution does have considerable width and tails, with a standard deviation (SD) of $17.53\mathrm{~km~s^{-1}~Mpc^{-1}}$. 

\begin{figure}[h!]
    \centering
    \includegraphics[width=\linewidth]{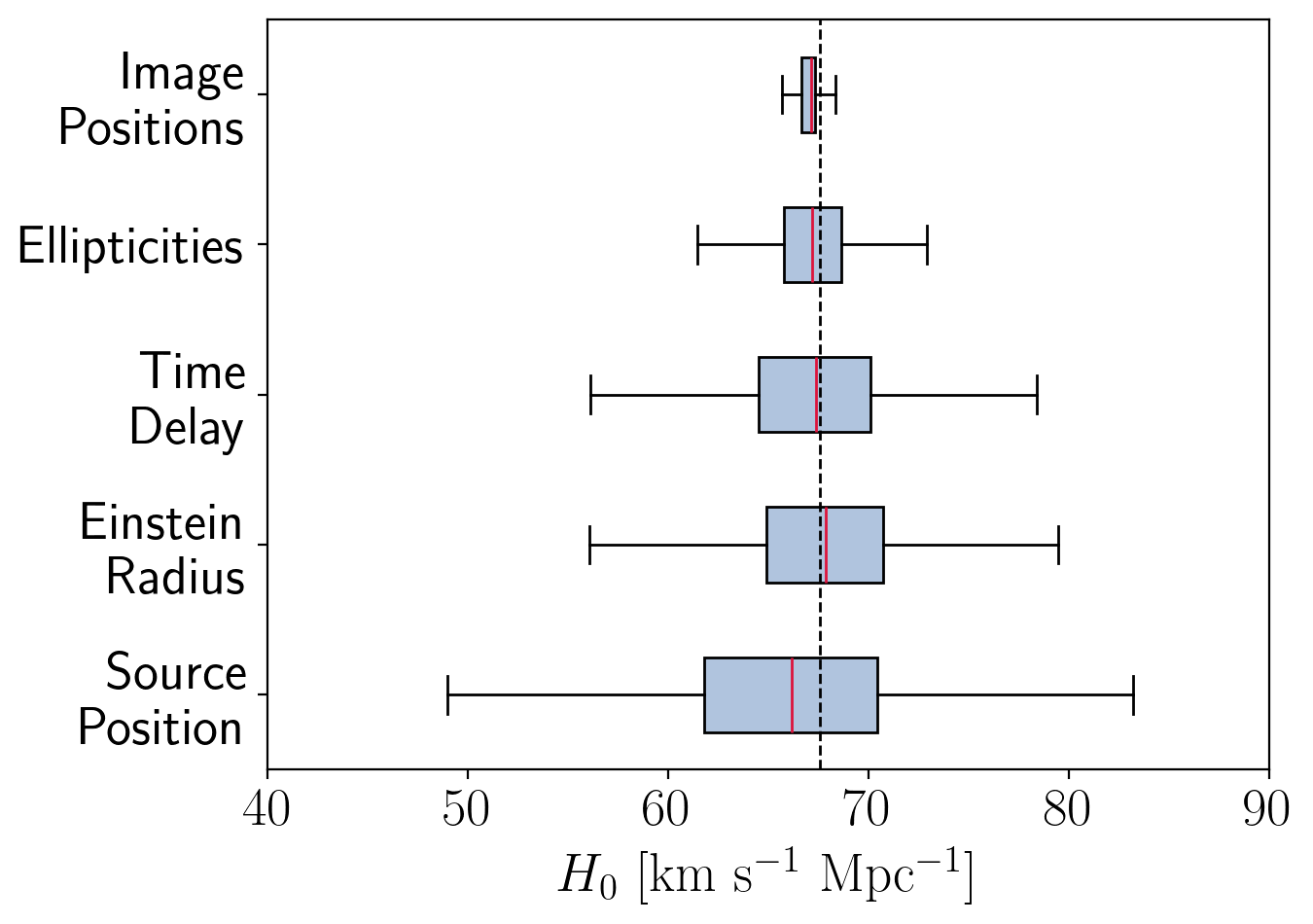}
    \caption{The contribution of each of the ensemble's models, on the overall scatter in the Hubble constant predictions. The blue shaded boxes represent the interquartile range (central 50\% of the data), the whiskers extend to 1.5 times the interquartile range from the first and third quartiles and the vertical red line represents the median of each distribution. The vertical dotted black line indicates the simulation's ground truth value of $H_0 = 67.7\mathrm{~km~s^{-1}~Mpc^{-1}}$. From top to bottom, taking only the corresponding model's output while fixing the remaining parameters to their ground truth values resulted in the following  $H_0$ median and SD values: \textbf{(i)} Image Positions: $67.15 \pm 4.8\mathrm{~km~s^{-1}~Mpc^{-1}}$;
      \textbf{(ii)} Ellipticities: $67.18 \pm 2.98\mathrm{~km~s^{-1}~Mpc^{-1}}$;
      \textbf{(iii)} Time Delay: $67.37 \pm 14.44\mathrm{~km~s^{-1}~Mpc^{-1}}$;
      \textbf{(iv)} Einstein Radius: $67.86 \pm 6.27\mathrm{~km~s^{-1}~Mpc^{-1}}$;
      \textbf{(v)} Source Position: $66.19 \pm 10.77\mathrm{~km~s^{-1}~Mpc^{-1}}$.}
    \label{fig:ensemble_contribution}
\end{figure}

Since an ensemble of CNN models was used, we can probe each model separately to assess its contribution towards the final prediction's dispersion. Instead of combining the predictions of all models for a single simulated glSN Ia, only the prediction for a single parameter (e.g., $\theta_E$) is used, and for the remaining parameters, the ground truth value is used. This is shown in Fig.~\ref{fig:ensemble_contribution}, where it can be seen that the most significant contribution to the estimate's uncertainty comes from the time delay and source position estimates. This demonstrates that it is not always the parameters with the largest relative ME, such as the ellipticity, that dominate the scatter in the final predictions. This useful capacity of ensembles to enable further investigation of each separate network enhances the network's interpretability and allows for more informed decisions on improving the ensemble's performance systematically.

\begin{figure*}
    \centering
    \includegraphics[width=\linewidth]{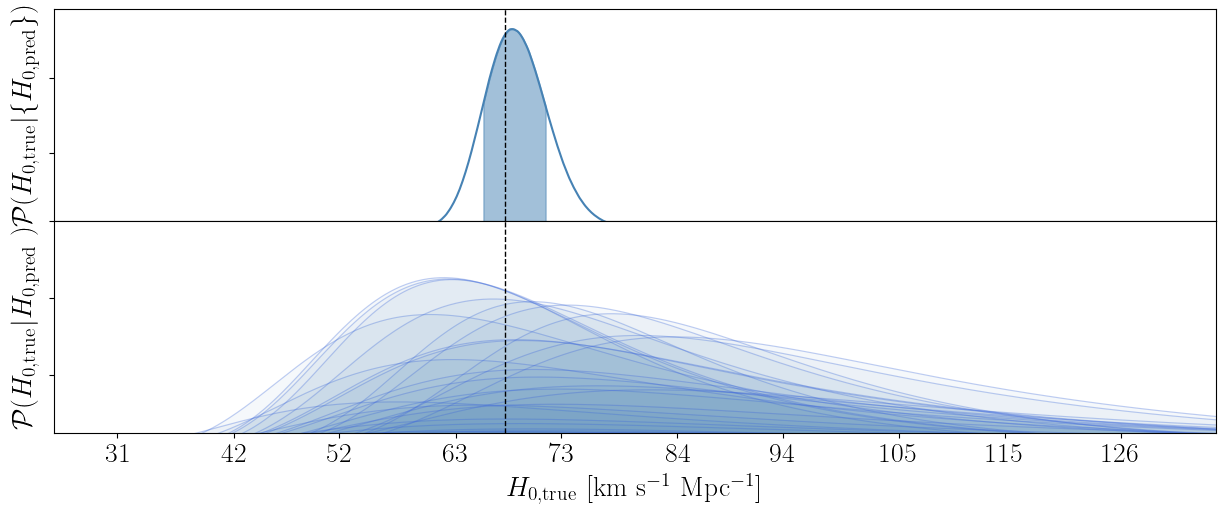}
    \caption{Joint $H_0$ posterior resulting from combining estimates from 100 independent glSNe Ia. In the upper panel, the combined posterior is shown, resulting in an estimate of $H_0 = \mathrm{(~68.61 \pm 3.05~) ~km~s^{-1}~Mpc^{-1}}$. The blue shaded region corresponds to the $1 \sigma$ region of the combined posterior, and the vertical black dashed line indicates the simulation's ground truth value, $H_0 = 67.7~\mathrm{km~s^{-1}~Mpc^{-1}}$.
    The lower panel displays all the individual posteriors that were combined for this estimate.}
    \label{fig:post_100}
\end{figure*}

\subsection{Obtaining posterior distributions}

To extend beyond the point estimates previously obtained, we employ the SBI procedure introduced in Sec.~\ref{subsect:SBI} to derive a posterior distribution for each network prediction. This methodology allows for a flexible combination of posterior distributions from multiple, independent glSNe Ia, resulting in a joint $H_0$ measurement. 
Motivated by the predicted glSN Ia rates as described in Sec.~\ref{subsect:Roman_telescope}, we combine the $H_0$ estimates for 100 systems. The resulting joint Hubble constant estimate from these systems is shown in Fig.~\ref{fig:post_100} and yields $H_0=\mathrm{(69.20 \pm 3.03) ~km~s^{-1}~Mpc^{-1}}$, which is within $1 \sigma$ of the simulation's ground truth value of $67.7~\mathrm{km~s^{-1}~Mpc^{-1}}$. 
The $4.4\%$ Hubble constant measurement presented here is a conservative estimate meant to demonstrate the capabilities of our method, since we are only including Roman Space Telescope data in one band and no follow-up observations.
Our results highlight that not only is the ensemble model capable of combining several $H_0$ estimates into a more precise measurement including robust uncertainties, but it also constitutes a fast and automated approach for achieving this. Predictions can be made on a daily basis as new data is acquired, and the knowledge about each parameter's contribution to the final $H_0$ uncertainty is particularly useful in defining strategies for follow-up observations.

\section{Discussion and conclusion}\label{Sec:Discussion}

We have designed a 3D CNN ensemble model, implemented within an SBI framework, to infer $H_0$ posteriors from an image time-series of glSN Ia. Our work is the first to combine ML-based estimates of the lens parameters with their corresponding time delays, in order to yield a measurement of the Hubble constant. Specifically, our model is trained and applied to simulated photometric images and their time evolution from the Roman Space Telescope.

We employed SBI to obtain full posterior distributions for each network prediction. The joint estimate resulting from the combination of 100 glSNe Ia posteriors was $H_0 = (69.20 \pm 3.03)~\mathrm{km~s^{-1}~Mpc^{-1}}$, which is consistent up to one SD with the ground truth of the simulation.
We emphasize that this $4.4\%$ measurement of the Hubble constant is a conservative estimate, intended to showcase the capabilities of our method, as it is based solely on Roman Space Telescope data in a single band and excludes any follow-up observations.
Furthermore, we demonstrated how the ensemble structure can be used to analyze each 3D CNN individually, where each network corresponds to predictions for a single lens or time-delay parameter, to evaluate their impact on the uncertainty in the final $H_0$ estimate. Through this analysis, we identified the time delay and source position estimates as the primary contributors to the overall uncertainty.

Several assumptions and design choices were made in this work, where the main focus was to demonstrate the capabilities of the ML models. Future studies can expand on this work through several possible approaches outlined below, particularly by increasing the fidelity of our glSN simulations.

Firstly, we limited ourselves to $F087$-band data from the Roman Space Telescope, but our network architecture can be modified to have multiple channels as input, each corresponding to a different band. Including the other $7$ bands from the Roman Space Telescope will considerably improve the cadence, and consequently, the precision on the time-delay estimates and the Hubble constant. Additional data from other telescopes could also be used, both from observations with an additional cadence (e.g., the Vera C. Rubin observatory) or with higher image resolution (e.g., the James Webb Space Telescope). 

Secondly, we only considered type Ia supernovae, although the number of expected core-collapse glSNe is even greater \citep{wojtak2019magnified, goldstein2019rates, oguri2010gravitationally}.  Further research is required to explore the cosmological prospects of core-collapse glSNe, which exhibits greater intrinsic variation than SNe Ia. Our simulated dataset could be further improved by including quadruply imaged glSNe Ia and microlensing effects from compact objects in the lens galaxy. We could also impose tighter quality cuts on the characteristics of the simulated systems to create a subset of glSNe Ia more suitable for cosmological inference (e.g. with longer time delays and larger image separations).

Thirdly, the lens mass distribution in our simulations was modeled using the SIE model, due to its simplicity and reasonable approximation of elliptical galaxy mass profiles. However, it has been shown \citep{sonnenfeld2018profiles, kochanek2020overconstrained, tdcosmo} that it can lead to over-constrained mass profiles and can possibly bias the estimates of the radial density profile, and consequently, $H_0$. As further investigation for this work, more complex lens models or non-parametric approaches to describe the lens potential could be considered.

An additional avenue to make our simulated difference images more realistic is by performing a manual subtraction of a reference image, instead of assuming a perfect subtraction of the lens and host galaxy. This would include some image subtraction artifacts into our simulated dataset, thereby making them more similar to real transient survey images. Training the network on such data will render it more robust to image subtraction artifacts in real data.

Finally, looking at how each ensemble model contributes to the final dispersion of $H_0$ estimates (c.f. Fig.~\ref{fig:ensemble_contribution}), it becomes apparent that improving the source position and time-delay predictions would be the most beneficial.

Overall, this work showcases how to build an ML pipeline to obtain robust $H_0$ estimates, which can be quickly deployed as more data is gathered. Furthermore, by employing an ensemble of models, as opposed to a single larger model, the parameters that most contribute to the error can be identified, informing how to improve the ensemble's performance systematically. With the upcoming order-of-magnitude increase of glSN detections from future surveys, this pipeline is a valuable tool for obtaining fast $H_0$ estimates and informing follow-up strategies.

\section*{Acknowledgments}

We would like to thank William D’Arcy Kenworthy for insightful discussions about \texttt{SALT3}, Jacob Osman Hjortlund for his valuable help with SBI, and Satadru Bag for his thoughtful feedback. Finally, Gonçalo Gonçalves thanks the lensed supernova group at the Oskar Klein Centre for their support, insightful discussions, and for providing a welcoming and enriching research environment. Rados{\l}aw Wojtak was supported by a research grant (VIL54489) from VILLUM FONDEN.

This work has been enabled by support from the research project grant ‘Understanding the Dynamic Universe’ funded by the Knut and Alice Wallenberg Foundation under Dnr KAW 2018.0067.
\\

\medskip
\textit{Software used:} Astropy \citep{Astropy2013, Astropy2018, Astropy2022},
Jupyter \citep{JupyterNotebook},
Matplotlib \citep{Matplotlib2007, Matplotlib2020},
NumPy \citep{Numpy202},
Pandas \citep{Pandas2010, Pandas2023},
Pickle \citep{Pickle2020},
SciPy \citep{Scipy2020},
Lenstronomy \citep{lenstronomy, Birrer2021LenstronomyII},
SNCosmo \citep{sncosmo},
PyTorch \citep{pytorch}

\section*{Data Availability}

The code for our glSN simulations and ML models is publicly available in this
\href{https://github.com/G-GoncaIves/Hubble_Constant_Inference}{GitHub Repository}.

\bibliographystyle{mnras}
\bibliography{references.bib}

\newcommand{\noop}[1]{}
\begin{thebibliography}{}
\makeatletter
\relax
\def\mn@urlcharsother{\let\do\@makeother \do\$\do\&\do\#\do\^\do\_\do\%\do\~}
\def\mn@doi{\begingroup\mn@urlcharsother \@ifnextchar [ {\mn@doi@} {\mn@doi@[]}}
\def\mn@doi@[#1]#2{\def\@tempa{#1}\ifx\@tempa\@empty \href {http://dx.doi.org/#2} {doi:#2}\else \href {http://dx.doi.org/#2} {#1}\fi \endgroup}
\def\mn@eprint#1#2{\mn@eprint@#1:#2::\@nil}
\def\mn@eprint@arXiv#1{\href {http://arxiv.org/abs/#1} {{\tt arXiv:#1}}}
\def\mn@eprint@dblp#1{\href {http://dblp.uni-trier.de/rec/bibtex/#1.xml} {dblp:#1}}
\def\mn@eprint@#1:#2:#3:#4\@nil{\def\@tempa {#1}\def\@tempb {#2}\def\@tempc {#3}\ifx \@tempc \@empty \let \@tempc \@tempb \let \@tempb \@tempa \fi \ifx \@tempb \@empty \def\@tempb {arXiv}\fi \@ifundefined {mn@eprint@\@tempb}{\@tempb:\@tempc}{\expandafter \expandafter \csname mn@eprint@\@tempb\endcsname \expandafter{\@tempc}}}

\bibitem[\protect\citeauthoryear{{Arendse} et~al.,}{{Arendse} et~al.}{2024}]{44_sne}
{Arendse} N.,  et~al., 2024, \mn@doi [\mnras] {10.1093/mnras/stae1356}, \href {https://ui.adsabs.harvard.edu/abs/2024MNRAS.531.3509A} {531, 3509}

\bibitem[\protect\citeauthoryear{{Arendse} et~al.,}{{Arendse} et~al.}{2025}]{Arendse2025}
{Arendse} N.,  et~al., 2025, \mn@doi [arXiv e-prints] {10.48550/arXiv.2501.01578}, \href {https://ui.adsabs.harvard.edu/abs/2025arXiv250101578A} {p. arXiv:2501.01578}

\bibitem[\protect\citeauthoryear{{Astropy Collaboration} et~al.,}{{Astropy Collaboration} et~al.}{2013}]{Astropy2013}
{Astropy Collaboration} et~al., 2013, \mn@doi [\aap] {10.1051/0004-6361/201322068}, \href {https://ui.adsabs.harvard.edu/abs/2013A&A...558A..33A} {558, A33}

\bibitem[\protect\citeauthoryear{{Astropy Collaboration} et~al.,}{{Astropy Collaboration} et~al.}{2018}]{Astropy2018}
{Astropy Collaboration} et~al., 2018, \mn@doi [\aj] {10.3847/1538-3881/aabc4f}, \href {https://ui.adsabs.harvard.edu/abs/2018AJ....156..123A} {156, 123}

\bibitem[\protect\citeauthoryear{{Astropy Collaboration} et~al.,}{{Astropy Collaboration} et~al.}{2022}]{Astropy2022}
{Astropy Collaboration} et~al., 2022, \mn@doi [\apj] {10.3847/1538-4357/ac7c74}, \href {https://ui.adsabs.harvard.edu/abs/2022ApJ...935..167A} {935, 167}

\bibitem[\protect\citeauthoryear{{Auger}, {Treu}, {Bolton}, {Gavazzi}, {Koopmans}, {Marshall}, {Bundy}  \& {Moustakas}}{{Auger} et~al.}{2009}]{Auger2009}
{Auger} M.~W.,  {Treu} T.,  {Bolton} A.~S.,  {Gavazzi} R.,  {Koopmans} L.~V.~E.,  {Marshall} P.~J.,  {Bundy} K.,   {Moustakas} L.~A.,  2009, \mn@doi [\apj] {10.1088/0004-637X/705/2/1099}, \href {https://ui.adsabs.harvard.edu/abs/2009ApJ...705.1099A} {705, 1099}

\bibitem[\protect\citeauthoryear{{Avestruz}, {Li}, {Zhu}, {Lightman}, {Collett}  \& {Luo}}{{Avestruz} et~al.}{2019}]{avestruz2019automated}
{Avestruz} C.,  {Li} N.,  {Zhu} H.,  {Lightman} M.,  {Collett} T.~E.,   {Luo} W.,  2019, \mn@doi [\apj] {10.3847/1538-4357/ab16d9}, \href {https://ui.adsabs.harvard.edu/abs/2019ApJ...877...58A} {877, 58}

\bibitem[\protect\citeauthoryear{Barbary et~al.,}{Barbary et~al.}{2025}]{sncosmo}
Barbary K.,  et~al., 2025, SNCosmo, \mn@doi{10.5281/zenodo.14714968}, \url {https://doi.org/10.5281/zenodo.14714968}

\bibitem[\protect\citeauthoryear{{Birrer} \& {Amara}}{{Birrer} \& {Amara}}{2018}]{lenstronomy}
{Birrer} S.,  {Amara} A.,  2018, \mn@doi [Physics of the Dark Universe] {10.1016/j.dark.2018.11.002}, \href {https://ui.adsabs.harvard.edu/abs/2018PDU....22..189B} {22, 189}

\bibitem[\protect\citeauthoryear{{Birrer} \& {Treu}}{{Birrer} \& {Treu}}{2021}]{Birrer&Treu2021}
{Birrer} S.,  {Treu} T.,  2021, \mn@doi [\aap] {10.1051/0004-6361/202039179}, \href {https://ui.adsabs.harvard.edu/abs/2021A&A...649A..61B} {649, A61}

\bibitem[\protect\citeauthoryear{{Birrer, S.} et~al.,}{{Birrer, S.} et~al.}{2020}]{tdcosmo}
{Birrer, S.} et~al., 2020, \mn@doi [A&A] {10.1051/0004-6361/202038861}, 643, A165

\bibitem[\protect\citeauthoryear{{Birrer}, {Dhawan}  \& {Shajib}}{{Birrer} et~al.}{2021a}]{birrer2021standardizable}
{Birrer} S.,  {Dhawan} S.,   {Shajib} A.~J.,  2021a, arXiv e-prints, \href {https://ui.adsabs.harvard.edu/abs/2021arXiv210712385B} {p. arXiv:2107.12385}

\bibitem[\protect\citeauthoryear{{Birrer} et~al.,}{{Birrer} et~al.}{2021b}]{Birrer2021LenstronomyII}
{Birrer} S.,  et~al., 2021b, \mn@doi [J.\ Open Source Software] {10.21105/joss.03283}, \href {https://ui.adsabs.harvard.edu/abs/2021JOSS....6.3283B} {6, 3283}

\bibitem[\protect\citeauthoryear{{Birrer} et~al.,}{{Birrer} et~al.}{2024}]{Birrer2024}
{Birrer} S.,  et~al., 2024, \mn@doi [\ssr] {10.1007/s11214-024-01079-w}, \href {https://ui.adsabs.harvard.edu/abs/2024SSRv..220...48B} {220, 48}

\bibitem[\protect\citeauthoryear{{Ca{\~n}ameras} et~al.,}{{Ca{\~n}ameras} et~al.}{2020}]{canameras2020holismokesII}
{Ca{\~n}ameras} R.,  et~al., 2020, \mn@doi [\aap] {10.1051/0004-6361/202038219}, \href {https://ui.adsabs.harvard.edu/abs/2020A&A...644A.163C} {644, A163}

\bibitem[\protect\citeauthoryear{Caswell et~al.,}{Caswell et~al.}{2020}]{Matplotlib2020}
Caswell T.~A.,  et~al., 2020, matplotlib/matplotlib: REL: v3.3.2

\bibitem[\protect\citeauthoryear{{Chen} et~al.,}{{Chen} et~al.}{2022}]{sn_chen22}
{Chen} W.,  et~al., 2022, Transient Name Server Discovery Report, \href {https://ui.adsabs.harvard.edu/abs/2022TNSTR2196....1C} {2022-2196, 1}

\bibitem[\protect\citeauthoryear{{Cheng}, {Li}, {Conselice}, {Arag{\'o}n-Salamanca}, {Dye}  \& {Metcalf}}{{Cheng} et~al.}{2020}]{cheng2020identifying}
{Cheng} T.-Y.,  {Li} N.,  {Conselice} C.~J.,  {Arag{\'o}n-Salamanca} A.,  {Dye} S.,   {Metcalf} R.~B.,  2020, \mn@doi [\mnras] {10.1093/mnras/staa1015}, \href {https://ui.adsabs.harvard.edu/abs/2020MNRAS.494.3750C} {494, 3750}

\bibitem[\protect\citeauthoryear{{Cranmer}, {Brehmer}  \& {Louppe}}{{Cranmer} et~al.}{2019}]{cranmer2019frontier}
{Cranmer} K.,  {Brehmer} J.,   {Louppe} G.,  2019, preprint, \href {https://ui.adsabs.harvard.edu/abs/2019arXiv191101429C} {} (\mn@eprint {arXiv} {1911.01429})

\bibitem[\protect\citeauthoryear{{Davies}, {Serjeant}  \& {Bromley}}{{Davies} et~al.}{2019}]{davies2019using}
{Davies} A.,  {Serjeant} S.,   {Bromley} J.~M.,  2019, \mn@doi [\mnras] {10.1093/mnras/stz1288}, \href {https://ui.adsabs.harvard.edu/abs/2019MNRAS.487.5263D} {487, 5263}

\bibitem[\protect\citeauthoryear{{Diego}, {Bernstein}, {Chen}, {Goobar}, {Johansson}, {Kelly}, {M{\"o}rtsell}  \& {Nightingale}}{{Diego} et~al.}{2022}]{Diego2022}
{Diego} J.~M.,  {Bernstein} G.,  {Chen} W.,  {Goobar} A.,  {Johansson} J.~P.,  {Kelly} P.~L.,  {M{\"o}rtsell} E.,   {Nightingale} J.~W.,  2022, \mn@doi [\aap] {10.1051/0004-6361/202143009}, \href {https://ui.adsabs.harvard.edu/abs/2022A&A...662A..34D} {662, A34}

\bibitem[\protect\citeauthoryear{{Ding}, {Liao}, {Birrer}, {Shajib}, {Treu}  \& {Yang}}{{Ding} et~al.}{2021}]{ding2021improved}
{Ding} X.,  {Liao} K.,  {Birrer} S.,  {Shajib} A.~J.,  {Treu} T.,   {Yang} L.,  2021, preprint, \href {https://ui.adsabs.harvard.edu/abs/2021arXiv210308609D} {} (\mn@eprint {arXiv} {2103.08609})

\bibitem[\protect\citeauthoryear{{Dobler} \& {Keeton}}{{Dobler} \& {Keeton}}{2006}]{dobler2006microlensing}
{Dobler} G.,  {Keeton} C.~R.,  2006, \mn@doi [\apj] {10.1086/508769}, \href {https://ui.adsabs.harvard.edu/abs/2006ApJ...653.1391D} {653, 1391}

\bibitem[\protect\citeauthoryear{{Falco}, {Gorenstein}  \& {Shapiro}}{{Falco} et~al.}{1985}]{Falco1985}
{Falco} E.~E.,  {Gorenstein} M.~V.,   {Shapiro} I.~I.,  1985, \mn@doi [\apjl] {10.1086/184422}, \href {https://ui.adsabs.harvard.edu/abs/1985ApJ...289L...1F} {289, L1}

\bibitem[\protect\citeauthoryear{{Foxley-Marrable}, {Collett}, {Vernardos}, {Goldstein}  \& {Bacon}}{{Foxley-Marrable} et~al.}{2018}]{foxley2018standardization}
{Foxley-Marrable} M.,  {Collett} T.~E.,  {Vernardos} G.,  {Goldstein} D.~A.,   {Bacon} D.,  2018, \mn@doi [\mnras] {10.1093/mnras/sty1346}, \href {https://ui.adsabs.harvard.edu/abs/2018MNRAS.478.5081F} {478, 5081}

\bibitem[\protect\citeauthoryear{Freedman}{Freedman}{2021}]{cornegie-chicago}
Freedman W.~L.,  2021, \mn@doi [The Astrophysical Journal] {10.3847/1538-4357/ac0e95}, 919, 16

\bibitem[\protect\citeauthoryear{Frye et~al.,}{Frye et~al.}{2024}]{sn_hope}
Frye B.~L.,  et~al., 2024, \mn@doi [The Astrophysical Journal] {10.3847/1538-4357/ad1034}, 961, 171

\bibitem[\protect\citeauthoryear{{Gentile} et~al.,}{{Gentile} et~al.}{2021}]{gentile2021lenses}
{Gentile} F.,  et~al., 2021, preprint, \href {https://ui.adsabs.harvard.edu/abs/2021arXiv210505602G} {} (\mn@eprint {arXiv} {2105.05602})

\bibitem[\protect\citeauthoryear{{Germain}, {Gregor}, {Murray}  \& {Larochelle}}{{Germain} et~al.}{2015}]{germain2015made}
{Germain} M.,  {Gregor} K.,  {Murray} I.,   {Larochelle} H.,  2015, preprint, \href {https://ui.adsabs.harvard.edu/abs/2015arXiv150203509G} {} (\mn@eprint {arXiv} {1502.03509})

\bibitem[\protect\citeauthoryear{{Goldstein}, {Nugent}  \& {Goobar}}{{Goldstein} et~al.}{2019}]{goldstein2019rates}
{Goldstein} D.~A.,  {Nugent} P.~E.,   {Goobar} A.,  2019, \mn@doi [\apjs] {10.3847/1538-4365/ab1fe0}, \href {https://ui.adsabs.harvard.edu/abs/2019ApJS..243....6G} {243, 6}

\bibitem[\protect\citeauthoryear{{Goobar} et~al.,}{{Goobar} et~al.}{2017}]{sn_iptf}
{Goobar} A.,  et~al., 2017, \mn@doi [Science] {10.1126/science.aal2729}, \href {https://ui.adsabs.harvard.edu/abs/2017Sci...356..291G} {356, 291}

\bibitem[\protect\citeauthoryear{{Goobar} et~al.,}{{Goobar} et~al.}{2023}]{sn_zwicky}
{Goobar} A.,  et~al., 2023, \mn@doi [Nature Astronomy] {10.1038/s41550-023-01981-3}, \href {https://ui.adsabs.harvard.edu/abs/2023NatAs...7.1098G} {7, 1098}

\bibitem[\protect\citeauthoryear{Goodfellow, Bengio  \& Courville}{Goodfellow et~al.}{2016}]{goodfellow2016deep}
Goodfellow I.,  Bengio Y.,   Courville A.,  2016, Deep learning.
MIT press

\bibitem[\protect\citeauthoryear{{Gorenstein}, {Falco}  \& {Shapiro}}{{Gorenstein} et~al.}{1988}]{Gorenstein1988}
{Gorenstein} M.~V.,  {Falco} E.~E.,   {Shapiro} I.~I.,  1988, \mn@doi [\apj] {10.1086/166226}, \href {https://ui.adsabs.harvard.edu/abs/1988ApJ...327..693G} {327, 693}

\bibitem[\protect\citeauthoryear{{Guy} et~al.,}{{Guy} et~al.}{2007}]{guy2007salt2}
{Guy} J.,  et~al., 2007, \mn@doi [\aap] {10.1051/0004-6361:20066930}, \href {https://ui.adsabs.harvard.edu/abs/2007A&A...466...11G} {466, 11}

\bibitem[\protect\citeauthoryear{{Harris} et~al.,}{{Harris} et~al.}{2020}]{Numpy202}
{Harris} C.~R.,  et~al., 2020, \mn@doi [\nat] {10.1038/s41586-020-2649-2}, \href {https://ui.adsabs.harvard.edu/abs/2020Natur.585..357H} {585, 357}

\bibitem[\protect\citeauthoryear{{Hayes}, {Thorp}, {Mandel}, {Arendse}, {Grayling}  \& {Dhawan}}{{Hayes} et~al.}{2024}]{hayes2024}
{Hayes} E.~E.,  {Thorp} S.,  {Mandel} K.~S.,  {Arendse} N.,  {Grayling} M.,   {Dhawan} S.,  2024, \mn@doi [\mnras] {10.1093/mnras/stae1086}, \href {https://ui.adsabs.harvard.edu/abs/2024MNRAS.530.3942H} {530, 3942}

\bibitem[\protect\citeauthoryear{{Hezaveh}, {Perreault Levasseur}  \& {Marshall}}{{Hezaveh} et~al.}{2017}]{hezaveh2017fast}
{Hezaveh} Y.~D.,  {Perreault Levasseur} L.,   {Marshall} P.~J.,  2017, \mn@doi [\nat] {10.1038/nature23463}, \href {https://ui.adsabs.harvard.edu/abs/2017Natur.548..555H} {548, 555}

\bibitem[\protect\citeauthoryear{{Huang} et~al.,}{{Huang} et~al.}{2020a}]{huang2020discovering}
{Huang} X.,  et~al., 2020a, preprint, \href {https://ui.adsabs.harvard.edu/abs/2020arXiv200504730H} {} (\mn@eprint {arXiv} {2005.04730})

\bibitem[\protect\citeauthoryear{{Huang} et~al.,}{{Huang} et~al.}{2020b}]{huang2020finding}
{Huang} X.,  et~al., 2020b, \mn@doi [\apj] {10.3847/1538-4357/ab7ffb}, \href {https://ui.adsabs.harvard.edu/abs/2020ApJ...894...78H} {894, 78}

\bibitem[\protect\citeauthoryear{{Huber} \& {Suyu}}{{Huber} \& {Suyu}}{2024}]{Huber2024}
{Huber} S.,  {Suyu} S.~H.,  2024, \mn@doi [\aap] {10.1051/0004-6361/202449952}, \href {https://ui.adsabs.harvard.edu/abs/2024A&A...692A.132H} {692, A132}

\bibitem[\protect\citeauthoryear{{Huber} et~al.,}{{Huber} et~al.}{2021}]{huber2021timedelay}
{Huber} S.,  et~al., 2021, preprint, \href {https://ui.adsabs.harvard.edu/abs/2021arXiv210802789H} {} (\mn@eprint {arXiv} {2108.02789})

\bibitem[\protect\citeauthoryear{Hunter}{Hunter}{2007}]{Matplotlib2007}
Hunter J.~D.,  2007, \mn@doi [Computing in Science & Engineering] {10.1109/MCSE.2007.55}, 9, 90

\bibitem[\protect\citeauthoryear{{Ivezi{\'c}} et~al.,}{{Ivezi{\'c}} et~al.}{2019}]{ivezic2019lsst}
{Ivezi{\'c}} {\v{Z}}.,  et~al., 2019, \mn@doi [\apj] {10.3847/1538-4357/ab042c}, \href {https://ui.adsabs.harvard.edu/abs/2019ApJ...873..111I} {873, 111}

\bibitem[\protect\citeauthoryear{{Kelly} et~al.,}{{Kelly} et~al.}{2015}]{sn_refsdal}
{Kelly} P.~L.,  et~al., 2015, \mn@doi [Science] {10.1126/science.aaa3350}, \href {https://ui.adsabs.harvard.edu/abs/2015Sci...347.1123K} {347, 1123}

\bibitem[\protect\citeauthoryear{{Kelly} et~al.,}{{Kelly} et~al.}{2022}]{sn_kelly22}
{Kelly} P.,  et~al., 2022, Transient Name Server AstroNote, \href {https://ui.adsabs.harvard.edu/abs/2022TNSAN.169....1K} {169, 1}

\bibitem[\protect\citeauthoryear{{Kelly} et~al.,}{{Kelly} et~al.}{2023}]{Kelly2023_Refsdal_H0}
{Kelly} P.~L.,  et~al., 2023, \mn@doi [Science] {10.1126/science.abh1322}, \href {https://ui.adsabs.harvard.edu/abs/2023Sci...380.1322K} {380, abh1322}

\bibitem[\protect\citeauthoryear{{Kenworthy} et~al.,}{{Kenworthy} et~al.}{2021}]{salt3}
{Kenworthy} W.~D.,  et~al., 2021, \mn@doi [\apj] {10.3847/1538-4357/ac30d8}, \href {https://ui.adsabs.harvard.edu/abs/2021ApJ...923..265K} {923, 265}

\bibitem[\protect\citeauthoryear{{Kluyver} et~al.,}{{Kluyver} et~al.}{2016}]{JupyterNotebook}
{Kluyver} T.,  et~al., 2016, in , IOS Press.
pp 87--90, \mn@doi{10.3233/978-1-61499-649-1-87}

\bibitem[\protect\citeauthoryear{{Kochanek}}{{Kochanek}}{2020}]{kochanek2020overconstrained}
{Kochanek} C.~S.,  2020, \mn@doi [\mnras] {10.1093/mnras/staa344}, \href {https://ui.adsabs.harvard.edu/abs/2020MNRAS.493.1725K} {493, 1725}

\bibitem[\protect\citeauthoryear{{Kodi Ramanah}, {Arendse}  \& {Wojtak}}{{Kodi Ramanah} et~al.}{2021}]{ramanah2021AIdriven}
{Kodi Ramanah} D.,  {Arendse} N.,   {Wojtak} R.,  2021, preprint, \href {https://ui.adsabs.harvard.edu/abs/2021arXiv210712399K} {} (\mn@eprint {arXiv} {2107.12399})

\bibitem[\protect\citeauthoryear{{Kolatt} \& {Bartelmann}}{{Kolatt} \& {Bartelmann}}{1998}]{kolatt1998magnification}
{Kolatt} T.~S.,  {Bartelmann} M.,  1998, \mn@doi [\mnras] {10.1046/j.1365-8711.1998.01466.x}, \href {https://ui.adsabs.harvard.edu/abs/1998MNRAS.296..763K} {296, 763}

\bibitem[\protect\citeauthoryear{{Kormann}, {Schneider}  \& {Bartelmann}}{{Kormann} et~al.}{1994}]{sie}
{Kormann} R.,  {Schneider} P.,   {Bartelmann} M.,  1994, \aap, \href {https://ui.adsabs.harvard.edu/abs/1994A&A...284..285K} {284, 285}

\bibitem[\protect\citeauthoryear{{Lanusse}, {Ma}, {Li}, {Collett}, {Li}, {Ravanbakhsh}, {Mandelbaum}  \& {P{\'o}czos}}{{Lanusse} et~al.}{2018}]{lanusse2018deeplens}
{Lanusse} F.,  {Ma} Q.,  {Li} N.,  {Collett} T.~E.,  {Li} C.-L.,  {Ravanbakhsh} S.,  {Mandelbaum} R.,   {P{\'o}czos} B.,  2018, \mn@doi [\mnras] {10.1093/mnras/stx1665}, \href {https://ui.adsabs.harvard.edu/abs/2018MNRAS.473.3895L} {473, 3895}

\bibitem[\protect\citeauthoryear{LeCun, Bengio  et~al.}{LeCun et~al.}{1995}]{lecun1995convolutional}
LeCun Y.,  Bengio Y.,   et~al., 1995, The handbook of brain theory and neural networks, 3361, 1995

\bibitem[\protect\citeauthoryear{LeCun, Bottou, Bengio, Haffner  et~al.}{LeCun et~al.}{1998}]{lecun1998gradient}
LeCun Y.,  Bottou L.,  Bengio Y.,  Haffner P.,   et~al., 1998, Proceedings of the IEEE, 86, 2278

\bibitem[\protect\citeauthoryear{{Lecun}, {Bengio}  \& {Hinton}}{{Lecun} et~al.}{2015}]{lecun2015deep}
{Lecun} Y.,  {Bengio} Y.,   {Hinton} G.,  2015, \mn@doi [\nat] {10.1038/nature14539}, \href {http://adsabs.harvard.edu/abs/2015Natur.521..436L} {521, 436}

\bibitem[\protect\citeauthoryear{{M{\"o}rtsell}, {Johansson}, {Dhawan}, {Goobar}, {Amanullah}  \& {Goldstein}}{{M{\"o}rtsell} et~al.}{2020}]{Mortsell2020}
{M{\"o}rtsell} E.,  {Johansson} J.,  {Dhawan} S.,  {Goobar} A.,  {Amanullah} R.,   {Goldstein} D.~A.,  2020, \mn@doi [\mnras] {10.1093/mnras/staa1600}, \href {https://ui.adsabs.harvard.edu/abs/2020MNRAS.496.3270M} {496, 3270}

\bibitem[\protect\citeauthoryear{{Oguri} \& {Kawano}}{{Oguri} \& {Kawano}}{2003}]{oguri2003gravitational}
{Oguri} M.,  {Kawano} Y.,  2003, \mn@doi [\mnras] {10.1046/j.1365-8711.2003.06290.x}, \href {https://ui.adsabs.harvard.edu/abs/2003MNRAS.338L..25O} {338, L25}

\bibitem[\protect\citeauthoryear{{Oguri} \& {Marshall}}{{Oguri} \& {Marshall}}{2010}]{oguri2010gravitationally}
{Oguri} M.,  {Marshall} P.~J.,  2010, \mn@doi [\mnras] {10.1111/j.1365-2966.2010.16639.x}, \href {https://ui.adsabs.harvard.edu/abs/2010MNRAS.405.2579O} {405, 2579}

\bibitem[\protect\citeauthoryear{{Papamakarios}, {Pavlakou}  \& {Murray}}{{Papamakarios} et~al.}{2017}]{papamakarios2017MAF}
{Papamakarios} G.,  {Pavlakou} T.,   {Murray} I.,  2017, preprint, \href {https://ui.adsabs.harvard.edu/abs/2017arXiv170507057P} {} (\mn@eprint {arXiv} {1705.07057})

\bibitem[\protect\citeauthoryear{{Park}, {Wagner-Carena}, {Birrer}, {Marshall}, {Lin}, {Roodman}  \& {LSST Dark Energy Science Collaboration}}{{Park} et~al.}{2021}]{park2021largescale}
{Park} J.~W.,  {Wagner-Carena} S.,  {Birrer} S.,  {Marshall} P.~J.,  {Lin} J. Y.-Y.,  {Roodman} A.,   {LSST Dark Energy Science Collaboration} 2021, \mn@doi [\apj] {10.3847/1538-4357/abdfc4}, \href {https://ui.adsabs.harvard.edu/abs/2021ApJ...910...39P} {910, 39}

\bibitem[\protect\citeauthoryear{{Pascale} et~al.,}{{Pascale} et~al.}{2024}]{Pascale2024_H0pe_H0}
{Pascale} M.,  et~al., 2024, \mn@doi [arXiv e-prints] {10.48550/arXiv.2403.18902}, \href {https://ui.adsabs.harvard.edu/abs/2024arXiv240318902P} {p. arXiv:2403.18902}

\bibitem[\protect\citeauthoryear{{Paszke} et~al.,}{{Paszke} et~al.}{2019}]{pytorch}
{Paszke} A.,  et~al., 2019, \mn@doi [arXiv e-prints] {10.48550/arXiv.1912.01703}, \href {https://ui.adsabs.harvard.edu/abs/2019arXiv191201703P} {p. arXiv:1912.01703}

\bibitem[\protect\citeauthoryear{{Pierel}, {Rodney}, {Vernardos}, {Oguri}, {Kessler}  \& {Anguita}}{{Pierel} et~al.}{2020}]{pierel2020projected}
{Pierel} J.~D.~R.,  {Rodney} S.,  {Vernardos} G.,  {Oguri} M.,  {Kessler} R.,   {Anguita} T.,  2020, arXiv e-prints, \href {https://ui.adsabs.harvard.edu/abs/2020arXiv201012399P} {p. arXiv:2010.12399}

\bibitem[\protect\citeauthoryear{{Pierel}, {Rodney}, {Vernardos}, {Oguri}, {Kessler}  \& {Anguita}}{{Pierel} et~al.}{2021}]{sne_detection}
{Pierel} J.~D.~R.,  {Rodney} S.,  {Vernardos} G.,  {Oguri} M.,  {Kessler} R.,   {Anguita} T.,  2021, \mn@doi [\apj] {10.3847/1538-4357/abd8d3}, \href {https://ui.adsabs.harvard.edu/abs/2021ApJ...908..190P} {908, 190}

\bibitem[\protect\citeauthoryear{{Pierel} et~al.,}{{Pierel} et~al.}{2024}]{sn_encore}
{Pierel} J.~D.~R.,  et~al., 2024, \mn@doi [\apjl] {10.3847/2041-8213/ad4648}, \href {https://ui.adsabs.harvard.edu/abs/2024ApJ...967L..37P} {967, L37}

\bibitem[\protect\citeauthoryear{{Planck Collaboration} et~al.,}{{Planck Collaboration} et~al.}{2016}]{plank_h0}
{Planck Collaboration} et~al., 2016, \mn@doi [\aap] {10.1051/0004-6361/201525830}, \href {https://ui.adsabs.harvard.edu/abs/2016A&A...594A..13P} {594, A13}

\bibitem[\protect\citeauthoryear{{Planck Collaboration} et~al.,}{{Planck Collaboration} et~al.}{2020}]{plank}
{Planck Collaboration} et~al., 2020, \mn@doi [A&A] {10.1051/0004-6361/201833910}, 641, A6

\bibitem[\protect\citeauthoryear{{Refsdal}}{{Refsdal}}{1964}]{Refsdal}
{Refsdal} S.,  1964, \mn@doi [\mnras] {10.1093/mnras/128.4.307}, \href {https://ui.adsabs.harvard.edu/abs/1964MNRAS.128..307R} {128, 307}

\bibitem[\protect\citeauthoryear{Riess et~al.,}{Riess et~al.}{2022}]{shoes}
Riess A.~G.,  et~al., 2022, \mn@doi [The Astrophysical Journal Letters] {10.3847/2041-8213/ac5c5b}, 934, L7

\bibitem[\protect\citeauthoryear{{Rodney}, {Brammer}, {Pierel}, {Richard}, {Toft}, {O'Connor}, {Akhshik}  \& {Whitaker}}{{Rodney} et~al.}{2021}]{sn_requiem}
{Rodney} S.~A.,  {Brammer} G.~B.,  {Pierel} J. D.~R.,  {Richard} J.,  {Toft} S.,  {O'Connor} K.~F.,  {Akhshik} M.,   {Whitaker} K.~E.,  2021, \mn@doi [Nature Astronomy] {10.1038/s41550-021-01450-9}, \href {https://ui.adsabs.harvard.edu/abs/2021NatAs...5.1118R} {5, 1118}

\bibitem[\protect\citeauthoryear{{Saha}}{{Saha}}{2000}]{Saha2000}
{Saha} P.,  2000, \mn@doi [\aj] {10.1086/301581}, \href {https://ui.adsabs.harvard.edu/abs/2000AJ....120.1654S} {120, 1654}

\bibitem[\protect\citeauthoryear{{Sainz de Murieta}, {Collett}, {Magee}, {Weisenbach}, {Krawczyk}  \& {Enzi}}{{Sainz de Murieta} et~al.}{2023}]{SainzdeMurieta2023}
{Sainz de Murieta} A.,  {Collett} T.~E.,  {Magee} M.~R.,  {Weisenbach} L.,  {Krawczyk} C.~M.,   {Enzi} W.,  2023, \mn@doi [\mnras] {10.1093/mnras/stad3031}, \href {https://ui.adsabs.harvard.edu/abs/2023MNRAS.526.4296S} {526, 4296}

\bibitem[\protect\citeauthoryear{{Schaefer}, {Geiger}, {Kuntzer}  \& {Kneib}}{{Schaefer} et~al.}{2018}]{schaefer2018deep}
{Schaefer} C.,  {Geiger} M.,  {Kuntzer} T.,   {Kneib} J.~P.,  2018, \mn@doi [\aap] {10.1051/0004-6361/201731201}, \href {https://ui.adsabs.harvard.edu/abs/2018A&A...611A...2S} {611, A2}

\bibitem[\protect\citeauthoryear{{Schneider} \& {Sluse}}{{Schneider} \& {Sluse}}{2013}]{schneider2013MSD}
{Schneider} P.,  {Sluse} D.,  2013, \mn@doi [\aap] {10.1051/0004-6361/201321882}, \href {https://ui.adsabs.harvard.edu/abs/2013A&A...559A..37S} {559, A37}

\bibitem[\protect\citeauthoryear{{Schneider}, {Ehlers}  \& {Falco}}{{Schneider} et~al.}{1992}]{schneider1992lenses}
{Schneider} P.,  {Ehlers} J.,   {Falco} E.~E.,  1992, {Gravitational Lenses}.
Springer, \mn@doi{10.1007/978-3-662-03758-4}

\bibitem[\protect\citeauthoryear{{Schuldt}, {Suyu}, {Meinhardt}, {Leal-Taix{\'e}}, {Ca{\~n}ameras}, {Taubenberger}  \& {Halkola}}{{Schuldt} et~al.}{2020}]{schuldt2020efficient}
{Schuldt} S.,  {Suyu} S.~H.,  {Meinhardt} T.,  {Leal-Taix{\'e}} L.,  {Ca{\~n}ameras} R.,  {Taubenberger} S.,   {Halkola} A.,  2020, preprint, \href {https://ui.adsabs.harvard.edu/abs/2020arXiv201000602S} {} (\mn@eprint {arXiv} {2010.00602})

\bibitem[\protect\citeauthoryear{{Scolnic} \& {Kessler}}{{Scolnic} \& {Kessler}}{2016}]{xc_dist}
{Scolnic} D.,  {Kessler} R.,  2016, \mn@doi [\apjl] {10.3847/2041-8205/822/2/L35}, \href {https://ui.adsabs.harvard.edu/abs/2016ApJ...822L..35S} {822, L35}

\bibitem[\protect\citeauthoryear{{Shapiro}}{{Shapiro}}{1964}]{shapiro1964GR}
{Shapiro} I.~I.,  1964, \mn@doi [\prl] {10.1103/PhysRevLett.13.789}, \href {https://ui.adsabs.harvard.edu/abs/1964PhRvL..13..789S} {13, 789}

\bibitem[\protect\citeauthoryear{{Sonnenfeld}}{{Sonnenfeld}}{2018}]{sonnenfeld2018profiles}
{Sonnenfeld} A.,  2018, \mn@doi [\mnras] {10.1093/mnras/stx3105}, \href {https://ui.adsabs.harvard.edu/abs/2018MNRAS.474.4648S} {474, 4648}

\bibitem[\protect\citeauthoryear{{Spergel} et~al.,}{{Spergel} et~al.}{2015}]{spergel2015roman}
{Spergel} D.,  et~al., 2015, preprint, \href {https://ui.adsabs.harvard.edu/abs/2015arXiv150303757S} {} (\mn@eprint {arXiv} {1503.03757})

\bibitem[\protect\citeauthoryear{{Suyu}, {Marshall}, {Auger}, {Hilbert}, {Blandford}, {Koopmans}, {Fassnacht}  \& {Treu}}{{Suyu} et~al.}{2010}]{suyu2010dissecting}
{Suyu} S.~H.,  {Marshall} P.~J.,  {Auger} M.~W.,  {Hilbert} S.,  {Blandford} R.~D.,  {Koopmans} L.~V.~E.,  {Fassnacht} C.~D.,   {Treu} T.,  2010, \mn@doi [\apj] {10.1088/0004-637X/711/1/201}, \href {https://ui.adsabs.harvard.edu/abs/2010ApJ...711..201S} {711, 201}

\bibitem[\protect\citeauthoryear{{Suyu}, {Goobar}, {Collett}, {More}  \& {Vernardos}}{{Suyu} et~al.}{2024}]{Suyu2024}
{Suyu} S.~H.,  {Goobar} A.,  {Collett} T.,  {More} A.,   {Vernardos} G.,  2024, \mn@doi [\ssr] {10.1007/s11214-024-01044-7}, \href {https://ui.adsabs.harvard.edu/abs/2024SSRv..220...13S} {220, 13}

\bibitem[\protect\citeauthoryear{{Team}}{{Team}}{2023}]{Pandas2023}
{Team} T. P.~D.,  2023, {pandas-dev/pandas: Pandas}, Zenodo, \mn@doi{10.5281/zenodo.3509134}

\bibitem[\protect\citeauthoryear{{Treu} \& {Marshall}}{{Treu} \& {Marshall}}{2016}]{treu2016cosmography}
{Treu} T.,  {Marshall} P.~J.,  2016, \mn@doi [\aapr] {10.1007/s00159-016-0096-8}, \href {https://ui.adsabs.harvard.edu/abs/2016A&ARv..24...11T} {24, 11}

\bibitem[\protect\citeauthoryear{{Tripp}}{{Tripp}}{1998}]{tripp}
{Tripp} R.,  1998, \aap, \href {https://ui.adsabs.harvard.edu/abs/1998A&A...331..815T} {331, 815}

\bibitem[\protect\citeauthoryear{Van~Rossum}{Van~Rossum}{2020}]{Pickle2020}
Van~Rossum G.,  2020, The Python Library Reference, release 3.8.2.
Python Software Foundation

\bibitem[\protect\citeauthoryear{{Virtanen} et~al.,}{{Virtanen} et~al.}{2020}]{Scipy2020}
{Virtanen} P.,  et~al., 2020, \mn@doi [Nature Methods] {10.1038/s41592-019-0686-2}, \href {https://ui.adsabs.harvard.edu/abs/2020NatMe..17..261V} {17, 261}

\bibitem[\protect\citeauthoryear{{Wagner-Carena}, {Park}, {Birrer}, {Marshall}, {Roodman}, {Wechsler}  \& {LSST Dark Energy Science Collaboration}}{{Wagner-Carena} et~al.}{2021}]{wagnercarena2021hierarchical}
{Wagner-Carena} S.,  {Park} J.~W.,  {Birrer} S.,  {Marshall} P.~J.,  {Roodman} A.,  {Wechsler} R.~H.,   {LSST Dark Energy Science Collaboration} 2021, \mn@doi [\apj] {10.3847/1538-4357/abdf59}, \href {https://ui.adsabs.harvard.edu/abs/2021ApJ...909..187W} {909, 187}

\bibitem[\protect\citeauthoryear{{Weisenbach}, {Schechter}  \& {Pontula}}{{Weisenbach} et~al.}{2021}]{Weisenbach2021}
{Weisenbach} L.,  {Schechter} P.~L.,   {Pontula} S.,  2021, \mn@doi [\apj] {10.3847/1538-4357/ac2228}, \href {https://ui.adsabs.harvard.edu/abs/2021ApJ...922...70W} {922, 70}

\bibitem[\protect\citeauthoryear{{Weisenbach}, {Collett}, {de Murieta}, {Krawczyk}, {Vernardos}, {Enzi}  \& {Lundgren}}{{Weisenbach} et~al.}{2024}]{Weisenbach_2024}
{Weisenbach} L.,  {Collett} T.,  {de Murieta} A.~S.,  {Krawczyk} C.,  {Vernardos} G.,  {Enzi} W.,   {Lundgren} A.,  2024, \mn@doi [\mnras] {10.1093/mnras/stae1396}, \href {https://ui.adsabs.harvard.edu/abs/2024MNRAS.531.4349W} {531, 4349}

\bibitem[\protect\citeauthoryear{{W}es {M}c{K}inney}{{W}es {M}c{K}inney}{2010}]{Pandas2010}
{W}es {M}c{K}inney 2010, in {S}t\'efan van~der {W}alt {J}arrod {M}illman eds, {P}roceedings of the 9th {P}ython in {S}cience {C}onference. pp 56 -- 61, \mn@doi{10.25080/Majora-92bf1922-00a}

\bibitem[\protect\citeauthoryear{{Wojtak} \& {Hjorth}}{{Wojtak} \& {Hjorth}}{2024}]{2024MNRAS.533.2319W}
{Wojtak} R.,  {Hjorth} J.,  2024, \mn@doi [\mnras] {10.1093/mnras/stae1977}, \href {https://ui.adsabs.harvard.edu/abs/2024MNRAS.533.2319W} {533, 2319}

\bibitem[\protect\citeauthoryear{{Wojtak}, {Hjorth}  \& {Gall}}{{Wojtak} et~al.}{2019}]{wojtak2019magnified}
{Wojtak} R.,  {Hjorth} J.,   {Gall} C.,  2019, \mn@doi [\mnras] {10.1093/mnras/stz1516}, \href {https://ui.adsabs.harvard.edu/abs/2019MNRAS.487.3342W} {487, 3342}

\bibitem[\protect\citeauthoryear{Wong et~al.,}{Wong et~al.}{2019}]{h0licow}
Wong K.~C.,  et~al., 2019, \mn@doi [Monthly Notices of the Royal Astronomical Society] {10.1093/mnras/stz3094}, 498, 1420

\bibitem[\protect\citeauthoryear{{Yahalomi}, {Schechter}  \& {Wambsganss}}{{Yahalomi} et~al.}{2017}]{Yahalomi2017}
{Yahalomi} D.~A.,  {Schechter} P.~L.,   {Wambsganss} J.,  2017, preprint, \href {https://ui.adsabs.harvard.edu/abs/2017arXiv171107919Y} {} (\mn@eprint {arXiv} {1711.07919})

\makeatother
\end{thebibliography}

\end{document}